\documentclass[aps,onecolumn,superscriptaddress,floatfix]{revtex4}
 
\usepackage{bm}
\usepackage[colorlinks=true, citecolor=blue, urlcolor=blue]{hyperref}\usepackage{epsf,graphics,graphicx}
\usepackage{amsmath}
\usepackage{latexsym}
\usepackage{amssymb,amsmath,graphicx}
\usepackage{amsfonts}
\usepackage[caption=false]{subfig}
\usepackage{epsfig,latexsym,graphicx}
\usepackage{amssymb}
\usepackage{graphicx}

\newcommand{\beq}{\begin{equation}}
\newcommand{\eeq}{\end{equation}}

\usepackage{epsfig}

\begin{document}
\title{ Maximum Entropy Framework for a Universal  Rank Order distribution with Socio-economic Applications
}
\author{Abhik Ghosh}
\affiliation{Interdisciplinary Statistical Research Unit, Indian Statistical Institute, Kolkata 700108}
\email{abhianik@gmail.com}
\author{Preety Shreya}
\author{Banasri Basu}
\affiliation{Physics and Applied Mathematics Unit, Indian Statistical Institute, Kolkata 700108, India}
\email{sribbasu@gmail.com}

\date{\today}

\begin{abstract}
In this paper we derive the maximum entropy characteristics of a particular rank order  distribution, namely the discrete generalized beta distribution, 
which has recently been observed to be extremely useful in modelling many several rank-size distributions from different context in Arts and Sciences, 
as a two-parameter generalization of Zipf's law.
Although it has been seen to provide excellent fits for several real world empirical datasets, 
the underlying theory responsible for the success of this particular rank order distribution is not explored properly. 
Here we, for the first time, provide its generating process which describes it as a natural maximum entropy distribution  
under an  appropriate bivariate utility constraint. 
Further, considering the similarity of the proposed utility function with the usual logarithmic utility function from economic literature,
we have also explored its acceptability in universal modeling of different types of socio-economic factors within a country as well as across the countries. 
The  values of distributional parameters estimated through a rigorous statistical estimation method,  along with the $ entropy$ values, 
are used to characterize the distributions of all these socio-economic factors over the years. 
\end{abstract}
\maketitle

\section{Introduction}
\label{SEC:intro}
The idea that mathematics  and physics can illuminate human conduct dates back to the eighteenth century \cite{hume}. Later, it was proposed that 
there are general scientific laws describing human societies \cite{auguste}, and 
statistical analyses of human qualities began therefrom \cite{adolphe}. Today the sciences of sociophysics and econophysics draw on all these ideas in their attempts to explain the socio-economic behavior in societies. In modern use $social$ $physics$  refers to using $big$ $data$  analysis and the mathematical laws to understand the behavior of socio-economic patterns. The core idea is that the socio-economic data
contain mathematical patterns that are characteristic of how social interactions spread and their converge. 
With the availability of huge amount of empirical data for a abundance of measures of human interactions makes it possible to divulge the socio-economic patterns. Finding the mathematical shape of a distribution function is as simple as a curve fitting but the significance of the mathematical form used to fit it should be clarified and explained. It is also important to know if the shape of a given distribution function can be explained by an underlying generative principle.
Statistical physics plays a major role in this effort by bringing tools and concepts able to bridge theory and empirical results. 
In this respect, the principle of Maximum entropy (MaxEnt) from statistical physics \cite{jaynes} 
provides a natural tool for deriving many standard probability distributions \cite{kapurbook}. Principles underlying power-law distributions have been sought in various types of models. In the present paper our focus is to formulate a Maximum entropy framework for  a two-parameter Rank-Order (RO) distribution and  show that  
the MaxEnt principle used in estimating probabilistic models from appropriate constraints is also 
the underlying basis of many socio-economic models 
via this universal RO distribution \cite{Martinez-Mekler09, physica19, Ausloos/Cerqueti:2016, Alvarez-Martinez/etc:2011, Alvarez-Martinez/etc:2014, 
Oscar2017, Alvarez-Martinez/etc:2018}. Our next goal is to characterize the uncertainty over the time span in the respective distributions through entropy analysis and study their evolution.  

The rank order distributions are very useful in modeling several grouped or ungrouped socio-economic data 
where we can arrange ($rank$) the data-points or groups according to their importance ($ size $).
To model a variable based on $N$ data-points (or groups) arranged in decreasing order of $ importance$  
with $i$-th item having rank $r_i$ and size $n_i$,
the most commonly used RO distribution is the hyperbolic Pareto (Zipf's) law.
It assigns a probability for ranks ($r$) by the probability mass function 
$f_P(r) = A\cdot r^{-\nu}, 
$
for $r=1, \ldots, N$, where $A$ is the normalizing constant and $\nu>0$ is the scaling (exponent) parameter. 
This Pareto law has also seen to be the maximum entropy distribution under the restriction imposed 
in terms of the expected utility function, when we consider the popular logarithmic  utility,
$u(r)=\log(r)$, from standard economic literature \cite{wu}. 
However, as evident from several empirical applications, 
this Pareto or power law can yield a good fit to any empirical data-sets  only at the end of low-ranks (large sizes) \cite{kgbb,akb}
and fails miserably at the other end of the rank order distribution specially 
when the data exhibits a wider spectrum of rank-sizes \cite{physica19}.

As an alternative to the Pareto law, a recently developed RO distribution, 
namely the discrete generalized beta (DGB) distribution, proposed in  \cite{beta1,beta2}, 
is seen to provide much improved fits for a wider spectrum of rank size  data arising from 
different areas of arts and science \cite{Martinez-Mekler09}.
The probability mass function associated with the DGB distribution is defined in terms of 
two real scaling parameters $(a, b)$ and is given by 
\begin{equation}
f_{(a,b)}(r) = A\frac{(N+1-r)^b}{r^a}, ~~~~r=1, \ldots, N,
\label{EQ:RO}
\end{equation}  
with $A$ being the required normalizing constant. 
Several scientists  have
 shown its successful applications in modelling different types of rank ordered data \cite{Martinez-Mekler09, physica19, Ausloos/Cerqueti:2016, Alvarez-Martinez/etc:2011, Alvarez-Martinez/etc:2014, 
Oscar2017, Alvarez-Martinez/etc:2018}. 
These also include the study of  the social variable  like city sizes \cite{physica19} in discrete attempts
but no extensive  study is available considering different types of important socio-economic factors. 
Importantly, the underlying theory responsible for the success of this particular RO distribution is not explored properly.
This paper is the first to provide a derivation of this particularly useful RO distribution
as a natural maximum entropy distribution under the  appropriate bivariate utility constraints. 
Moreover, our analysis  will also illustrate its acceptability in universal modeling of different types of socio-economic factors. 
In particular, for validating our proposal we have considered data for diverse socio-economic  factors within a country as well as across the countries.  


The rest of the paper is organized as follows. Sec.~II  deals with the mathematical formulation of the Maximum Entropy Rank Order distribution . 
In Sec.~III we validate our theoretical proposal with empirical data sets of different socio-economic variables within a country for a large time span. The discussion on the uncertainty of the distribution is also narrated with the help of the entropy of the corresponding distribution. The corresponding subsections A, B and C respectively deals with city sizes of Japan, personal and household income of USA and election results of India for various years.
Furthermore, in Sec.IV we demonstrate our proposal with a few socio-economic indicators across different countries around the globe and do a comparative analysis in terms of entropy at different time points. The respective subsections A, B, and C of Sec~IV illustrates  the RO distribution and the corresponding Shannon entropies
for the population, percentage of agricultural land, GDP per capita and Buffet indicators of different countries across the world. 
Finally, we conclude in Sec.~V.

\section{The Maximum Entropy Rank Order Distributions}

Although initially introduced  for solving a problem in statistical mechanics \cite{jaynes}, 
the principle of Maximum Entropy has become a widely applied tool for constructing the probability distribution in statistical inference, 
decision science, pattern-recognition, communication theory, time-series analysis, and many more. 
The function which is maximized, namely the entropy, does have remarkable properties entitling it to be considered a good measure of the amount of uncertainty contained in a probability distribution \cite{pme}. 
We point out here that currently MaxEnt is being regarded as a principle that is broader than either physics or information alone.
It is also a procedure which ensures that any inference drawn from stochastic data satisfies the basic self-consistency requirements \cite{maxentrev}.


Let us consider an RO distribution having support $\{1, 2, \ldots, N\}$;
this can be used to model rank-size data on $N$ items (data-points). 
Suppose we use a model which assigns the probabilities to different ranks through a 
probability mass function (pmf) $f(r)$, $r=1, \ldots, N$. 
Note that, by definition, it satisfies 
\begin{eqnarray}
\sum_{r=1}^N f(r) = 1.
\label{EQ:Norm}
\end{eqnarray}

As the uncertainty in any model distribution can be measured by the Shannon-Gibb's entropy,  we use the distribution with maximum uncertainty 
(entropy) according to the Jaynes' MaxEnt principle \cite{jaynes}. 
The Shannon entropy of our model RO distribution $f(r)$ is given by
\begin{eqnarray}
S(f) &=& -\sum_{r=1}^N f(r) \log f(r),
\label{EQ:Entropy}
\end{eqnarray}
where the logarithm is taken with base $e$ (natural logarithm).
It is straightforward and intuitive to see that the pmf that maximizes the Shannon entropy in (\ref{EQ:Entropy})
subject to the normalizing constraints (\ref{EQ:Norm}) is the uniform distribution \cite{kapurbook} 
which assigns equal probability (of $\frac{1}{N}$) to each rank with the corresponding (global) maximum entropy value being $\log(N)$.
In practice, however, no data are uniformly distributed and cannot be modelled by the uniform distribution.
So, one generally add some more constraints into the optimization problem to get a meaningful class of
 model distribution
to be used to fit the empirical data. As a simple example, if we maximize the entropy of 
a random variable having support whole of the real line with 
the constraints of having known mean and variance,  we get the famous Gaussian (normal) distribution.  

In the contexts of socio-economic applications, the constraints to derive the 
maximum entropy distribution is often taken in terms of a known expected utility:
$E(u)=$ constant, for a given utility function $u$. The popular choice of this utility 
having enough economic justification is the logarithmic entropy
\begin{eqnarray}
u(r)=\log(r).
\label{EQ:utility1}
\end{eqnarray} 
For the RO distributions, if we maximize the the Shannon entropy in (\ref{EQ:Entropy})
subject to the expected utility constraint with the utility given in (\ref{EQ:utility1}),
i.e. $E(log(r))=$ constant,
along with the normalizing constraint (\ref{EQ:Norm}),
the resulting Max-Ent distribution turns out to the Pareto law.

Note that the utility used in the case of the Pareto distribution only restricts the end of lower ranks
by the consideration of the $\log(r)$ terms. Motivated by the failure in modeling the other end of ranks,
we propose to add one more constraint in terms of the terms $\log(N+1-r)$.
This leads to a bivariate utility function, under the rank-size context,  of the form  
\begin{eqnarray}
u(r)=\begin{pmatrix}
\begin{array}{c}
\log(r)\\
\log(N+1-r)
\end{array}
\end{pmatrix}.
\label{EQ:utility2}
\end{eqnarray} 
It is to be noted that the choice of this particular utility function puts equal weightage to both ends of the ranks 
keeping intact the interpretations and the justifications of logarithmic utility in socio-economic contexts.
We can then can consider the expected utility constraints in terms of the newly proposed bivariate utility function as
\begin{eqnarray}
E(\log(r))&=&\sum_{r=1}^N \log(r) f(r) = c_1,
\label{EQ:utility_constraints1}\\
E(\log(N+1-r))&=&\sum_{r=1}^N \log(N+1-r) f(r) =c_2,
\label{EQ:utility_constraints2}
\end{eqnarray} 
for two constants $c_1$ and $c_2$, and maximize the Shannon entropy in (\ref{EQ:Entropy})
under these conditions along with (\ref{EQ:Norm}). 
To solve this maximization problem, the method of Lagrange multipliers is used and as such  the objective function can be written as
\begin{eqnarray}
H(f_1, \ldots, f_N) &=& -\sum_{r=1}^N f_r \log f_r +\lambda_0\left(\sum_{r=1}^N f(r) - 1\right)
+\lambda_1\left(\sum_{r=1}^N \log(r)f_r -c_1\right) + \lambda_2\left(\sum_{r=1}^N \log(N+1-r)f_r -c_2\right),
\nonumber
\end{eqnarray}
where we denote $f_r=f(r)$ for $r=1, \ldots, N$ 
and $\lambda_0, \lambda_1, \lambda_2$ are the necessary  Lagrange multipliers. 
Taking derivatives with respect to $f_r$ lead to the equations
\begin{eqnarray}
-\log f_r - 1 +\lambda_0 +{\lambda_1}{\log r} + {\lambda_2}{\log(N+1-r)}=0, ~~~~~r=1, \ldots, N,
\label{EQ:EE1}
\end{eqnarray}
and derivatives with respect to $\lambda_0, \lambda_1, \lambda_2$ lead to 
Equations (\ref{EQ:Norm}), (\ref{EQ:utility_constraints1}) and (\ref{EQ:utility_constraints2}), respectively. 
Now, solving (\ref{EQ:EE1}), we get the Maximum entropy RO distribution $f_r^\ast$ as given by
\begin{eqnarray}
&&\log f_r^\ast =(\lambda_0-1) +{\log r^{\lambda_1}(N+1-r)^{\lambda_2}}=0, 
\nonumber\\
&& f_r^\ast = 2^{\lambda_0+1}r^{\lambda_1}(N+1-r)^{\lambda_2},~~~~~r=1, \ldots, N,
\end{eqnarray}
which is exactly in the same form as the DGB distribution in (\ref{EQ:RO})
with $a=-\lambda_1$, $b=\lambda_2$ and $A=2^{\lambda_0+1}$.
Note that $\lambda_0$ is to be obtained by normalizing the equations $\sum_{r=1}^N f_r^\ast = 1$
which makes equivalently $A$ as the normalizing constant.
The two other parameters $\lambda_1$ and $\lambda_2$, and equivalently $(a,b)$, 
depend on the given constant $c_1$ and $c_2$ and 
hence define a class of MaxEnt distributions given by (\ref{EQ:RO}).
The  scaling parameters $(a,b)$ also characterize the shape of the MaxEnt DGB distribution \cite{physica19}.
This analysis provides a natural way to obtain the DGB distribution as a maximum entropy distribution
under the expected utility constraints with an extremely intuitive bivariate utility function.

The corresponding value of the maximum Shannon entropy is given by 
\begin{eqnarray}\label{entropy1}
S_{\max}(a,b) = S(f^\ast) 
= -\log A -A\sum_{r=1}^N \frac{(N+1-R)^b}{r^a} [b \log (N+1-r)-a \log r],
\end{eqnarray}
which provides the maximum amount of uncertainly present in the corresponding RO distribution under the constraints in 
(\ref{EQ:utility_constraints1})--(\ref{EQ:utility_constraints2}) and hence depends on $c_1$, $c_2$ 
or equivalently on the distributional (scaling) parameters $(a,b)$; 
see \cite{physica19} for its behavior with changing parameter values.

Since the constants $c_1$ and $c_2$ in the Max-Ent derivation of the DGB distribution are not known in practice, 
while  using this MaxEnt distribution to model a real-life population, 
the associated parameters $(a,b)$ need to be based on the empirical data. 
We have used a simple, yet most efficient, method of estimation via the maximum likelihood approach
and subsequent model checking via the Kolmogorov-Smirnov (KS) error metric between 
the observed and the fitted (cumulative) probabilities; these methods are described briefly  in Appendix \ref{APP:A}.
Based on the estimated values $(\widehat{a}, \widehat{b})$ of the parameters $(a,b)$, 
we can then also estimate the maximum uncertainty present in the empirical distribution as 
$\widehat{S}_{\max} = S_{\max}(\widehat{a}, \widehat{b})$.
The estimated maximum entropy value $\widehat{S}_{\max}$ would help us to characterize the underlying socio-economic variable 
along with their distributional shape analyses via$(\widehat{a}, \widehat{b})$.

\section{Applications in Modeling Socio-economic factors within a single country }

In this section, we present the results obtained by the applications of the Maximum Entropy Rank-Order distribution for a few socio-economic variables within a single country. With the motivation to validate our result with different types of data which are diverse in character,
we have chosen city size, income data and election results for different time windows according to the availability of public data.    
The results of the goodness-of-fit test highly signify the importance of the proposed rank-order distribution for all such cases.  
Moreover, the entropy analysis explains the uncertainty in their distributions over the years in all cases.

\subsection{City Size distribution for Japan }

Despite being just the 61st largest country in the world by area, Japan is one of the most populous countries in the world. 
The last set of official figures \cite{japan}  pertaining to Japan’s population were released at the time of the 2015 census  
and the final statistics showed there were 127,094,745 people, which would make Japan the 11th largest country in the world. 
However, the most recent estimate places the number lower at 126.71 million, still the world's 10th most populous country. 
Though in decline, it still holds that position in 2019 with an estimated 126.85 million people. 
The data  show the population is shrinking in 39 areas of the country, and growing in eight. 
Japan's nine major urban areas account for 53.9 \% of the total population, with Greater Tokyo now home to 28.4\%. 
Rural areas, on the other hand, are being hit by severe declines. The official census of Japan is performed in every five years, 2015 being the latest.
According to latest estimate \cite{japan}, its largest city, Tokyo, has over 8 million residents 
while there are additional 13 cities with population over 1 Million.  
Besides, 191 cities have population in the range of 100,000 to 1 Million and  540 cities with number of people within the range of 10,000 to 100,000.  
Here, we consider the census data of the population of Japanese cities with at least 20000 inhabitants 
for the years 1995-2010.

\begin{figure}[h]
\subfloat[Estimates of parameters $(a,b)$]{
	\includegraphics[width=0.5\textwidth]{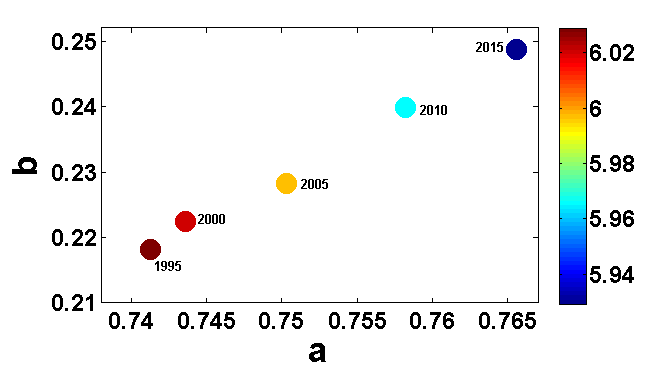}
		\label{FIG:Japan_ab}
}
\subfloat[Estimated Entropy $\widehat{S}_{\max}$]{
	\includegraphics[width=0.5\textwidth]{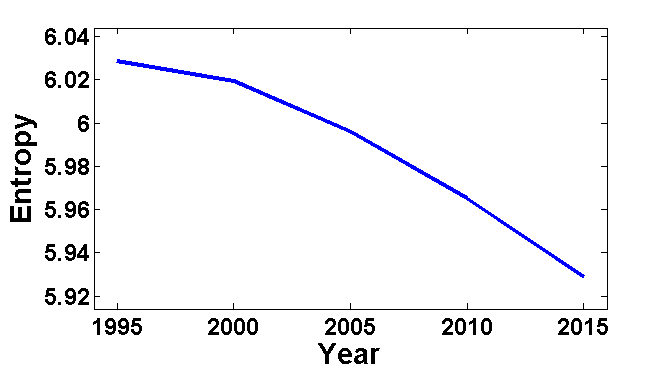}
	\label{FIG:Japan_Ent}
}
\caption{Estimates of the parameters $(a,b)$ and the maximum entropy values ($\widehat{S}_{\max}$)  for the MaxEnt RO distribution 
	fitted  to the Japan's city-size data  over different years} 
\label{FIG:Japan}
\end{figure}

\bigskip\bigskip
We have fitted the MaxEnt RO distribution, i.e. the DGB distribution, 
for the empirical data of Japan's city-sizes for each census years.
For brevity, detailed results are presented in Supplementary Table \ref{TAB:Japan} and Figure 8, 
which show the excellent fits for our DGD distribution for each years with the KS error measure being only 0.0133 -- 0.0149.
To illustrate the power of the proposed MaxEnt approach for further in-depth analyses, 
we present the estimates of the scaling parameters $(a, b)$ 
and the maximum entropy value $\widehat{S}_{\max}$ in Figure \ref{FIG:Japan}.
One can clearly observe  that the entropy of the Japan's city-size distribution is decreasing sharply over the years;
it signifies that the larger cities are becoming even larger and smaller cities are becoming even smaller in general.
Further, there is a consistent change in the shape of the fitted RO distribution with increasing values of both the scaling parameters $(a,b)$.
These provide a mathematical explanation, along with the underlying mechanism, of the fact that 
more people are continuously moving from rural areas or small towns in Japan to the larger cities over the  past twenty years.
Urban city dynamics of Japan is thus becoming vulnerably extreme and needs urgent planning for their sustainability.

\subsection{Income distributions of USA: Personal vs. Household Incomes}

As our second illustration, we consider grouped data on the personal and the household income in USA for the years 2012 to 2017.
These  data are collected  from the Current Population Survey (CPS), an annual social and economic supplement, 
sponsored jointly by the U.S. Census Bureau and the U.S. Bureau of Labor Statistics (BLS). 
The available grouped data \cite{usapnc,usahh} provide the number of people (in thousands) or households in 41 different income groups; 
the personal income refers to people of ages 15 years and above but the household income data covers \textit{all} households in USA.
From the years 2012 only, the income classes are given in the interval of \$2499 and \$4999, respectively, 
for the personal and household income data; 
the respective last groups correspond to the persons having income $\geq\$100000$ 
and the households having income is $\geq$ \$ 200000 per /year.

\begin{figure}[h]
	\subfloat[Estimates of $(a,b)$ for Household income]{
		\includegraphics[width=0.5\textwidth]{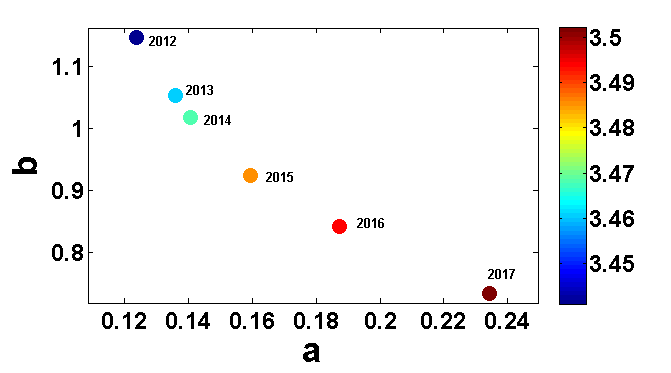}
		\label{FIG:Income_ab1}
	}
	\subfloat[Estimates of $(a,b)$ for Personal income ]{
	\includegraphics[width=0.5\textwidth]{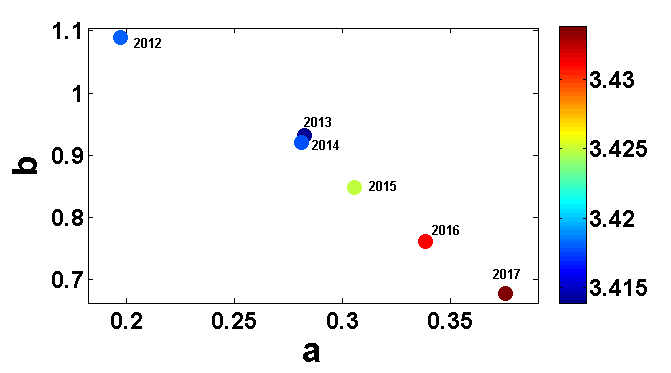}
	\label{FIG:Income_ab2}
}\\
	\subfloat[Estimated Entropies ($\widehat{S}_{\max}$)]{
		\includegraphics[width=0.55\textwidth]{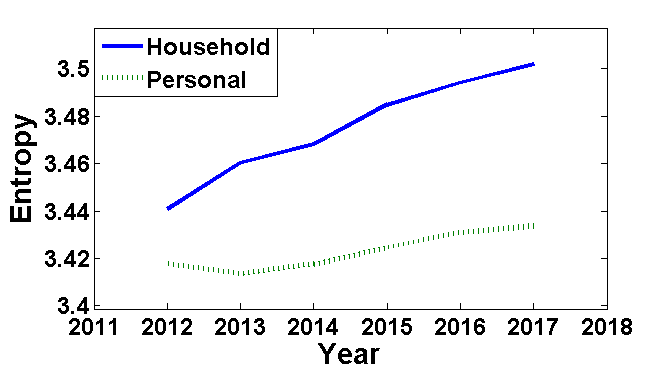}
		\label{FIG:Income_Ent}
	}
\caption{Estimates of the parameters $(a,b)$ and the maximum entropy values $\widehat{S}_{\max}$  for the MaxEnt RO distribution 
	fitted  to the personal and household income data of USA over different years} 
	\label{FIG:Income}
\end{figure}

We employ our MaxEnt approach to fit the DGB distribution to these data which again yields remarkably good fits
having KS error in the range [0.0149, 0.0255] only; the detailed results are given in Supplementary Tables \ref{TAB: USAP}-\ref{TAB: USAH} and Supplementary Figures 0-10,
and the parameter estimates  along with $\widehat{S}_{\max}$ are plotted in Figure \ref{FIG:Income} 
for both personal and household income data. Note that, for both the personal and household income, the scaling parameters 
follow a prominent changing pattern over the years --- values of $b$ are in a similar range in both cases and decrease over time,
whereas the value of $a$ is increasing over the years and are higher for personal income data. 
Further, in contrast to Japan's city-size distribution, 
the entropies $\widehat{S}_{\max}$ of the fitted MaxEnt distribution to both the personal and household incomes of USA
are seen to have an increasing trend over the years. 
These indicate that the difference in the number of people/households  in different income groups are decreasing in general.

The entropy and its increment rate are much greater for the household incomes compared to the personal incomes.
Thus, the  household income distributions in USA are relatively closer to the uniform distribution than personal incomes in each year
and further moving towards more uniformity  over the time.  Does it signify  that the income inequality among households in USA
are decreasing over time? Although it needs further focused investigations considering household sizes, 
there is definite hope since the entropy in 2017 is not very far from the global maximum value (corresponding to uniform distribution) 
of $\log(N)$ = 3.714.  For personal income, however, the rate of increment in significantly slower over the last five years 
and might look to stabilize in more recent years; the income inequality among people of USA is still looking  far from being close to uniformity.

\subsection{ Indian Lok Sabha Election Results} 

Not all socio-economic factors behave so nicely as in the previous two examples if they are not governed by a well-structured economy 
and hence also depends on the corresponding country. To illustrate the applicability of our proposed MaxEnt RO approach beyond such cases as well,
our third illustration here is to focus  on a completely different type of data on the winners of Indian Lok Sabha elections.
India is a federation with a parliamentary system governed under the Constitution of India, 
where the Parliament consists of the President of India and the two Houses of Parliament 
known as the Council of States (Rajya Sabha) and the House of the People (Lok Sabha).
Members of the Lok Sabha are elected by adult universal suffrage and a first-past-the-post  system to represent their respective constituencies;
more lucidly,  they are elected by being voted upon by all adult citizens of India, from a set of candidates who stand in their respective constituencies. 
Every adult citizen of India can vote only one candidate and only in their own constituency, 
and the candidate with the highest number of votes is elected.  
Although the maximum strength of the House allotted by the Constitution of India is 552, 
currently it has 545 members including 543 (at max) elected members and 2 (at max) nominated members of the Anglo-Indian Community by the President of India. 
The Lok Sabha (House of the People) was duly constituted for the first time on 17 April 1952 
after the first General Elections held from 25 October 1951 to 21 February 1952. 
Subsequently, following the same procedure and as the situation arose, later the election was conducted 17 times till date
most of which are in a five year interval (except few situations of political crisis). 
Due to the complex nature of the process, it is a definite challenge to mathematically model and analyze the voting pattern in different Lok Sabha 
elections and their evolution over time.
Here, for the first time, we model the winner's vote percentage and the percentage of vote-margin of the winning candidate from the second-placed candidate
for a Lok Sabha election in India by the proposed MaxEnt approach with the DGB distribution. 
As per the availability of the data \cite{election}, we consider the 11 Lok Sabha elections from the year 1980 to 2019.


\begin{figure}[h]
	\subfloat[Estimates of parameters $(a,b)$ for Winner's Vote \%]{
		\includegraphics[width=0.5\textwidth]{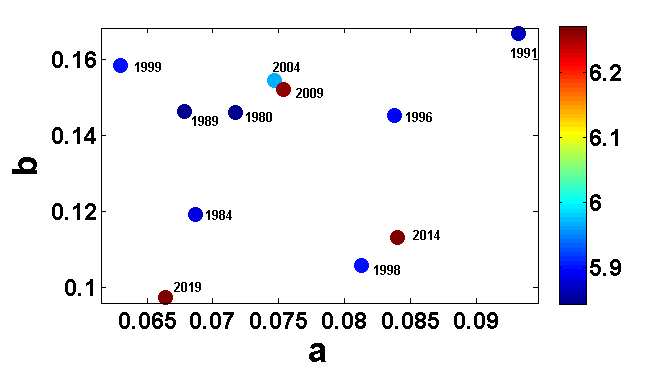}
		\label{FIG:ECindia_ab1}
	}
	\subfloat[Estimates of parameters $(a,b)$ for Winner's Vote-margin \%]{
		\includegraphics[width=0.5\textwidth]{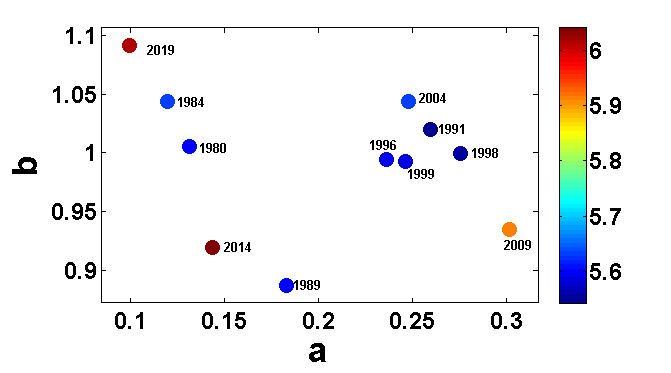}
		\label{FIG:ECindia_ab2}
	}\\
	\subfloat[Estimated Entropy $\widehat{S}_{\max}$]{
		\includegraphics[width=0.5\textwidth]{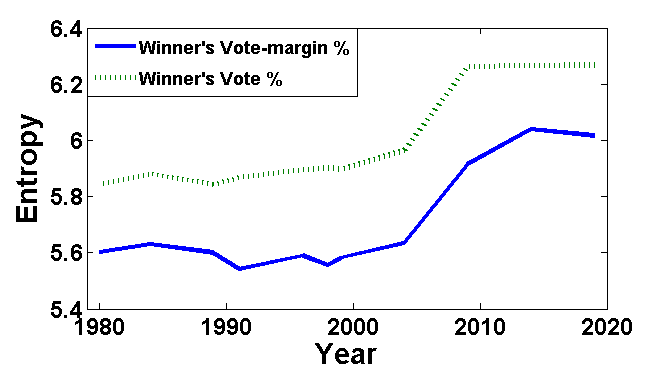}
		\label{FIG:ECindia_Ent}
	}
\caption{Estimates of the parameters $(a,b)$ and the maximum entropy values $\widehat{S}_{\max}$  for the MaxEnt RO distribution 
	fitted  to the data on Winner's vote percentage and Vote-margin percentage in Indian Lok Sabha elections in different years} 
	\label{FIG:E}
\end{figure}

It is really interesting to see the effectiveness of the MaxEnt DGB distribution to produce excellent fits for such a complex phenomena as well
with extremely low KS error in the range 0.0019 to 0.0084 only; see Supplementary Tables \ref{TAB:ECIWP}--\ref{TAB:ECIWMP} for detailed results 
and Supplementary Figures 11--12 for the actual fits. The parameter estimates and the maximum entropy values are plotted in Figure \ref{FIG:E}.
Although there is no clear pattern of the distributional shapes (i.e., estimated values of $a$ and $b$) over the years,
their entropy ($\widehat{S}_{\max}$) follow a clearly similar trend. 
However, the entropies of the winner's vote percentage are seen to be significantly higher than those of he vote-margin percentage;
the values of  $\widehat{S}_{\max}$ in the first case is almost the same as the corresponding global maximum ($\log(N)$)
so that the winner's vote percentage  is mostly uniform in every Indian Lok Sabha election. 
Further, for both the cases,  $\widehat{S}_{\max}$  has a significant increment (structural break) from the year 2004 to 2009,
after remaining mostly stable before 2004. Other than a significant increase in the number of constituencies ($N$),
there might also be some socio-political reasons to explain such a drastic change in voting pattern 
which will be an interesting future question to investigate.

  \section{Applications in Global Modeling across different countries}

In this section, we will illustrate the universality of the MaxEnt RO distribution 
even for global modeling of various socio-economic indicators across different countries. 
In particular, we have considered the data for  population, percentages of available agricultural land,  per capita GDP 
and the ratio of market capitalization to GDP (known as the \textit{Buffet indicator}) from different countries across the world;
the time frames of the analyses and the countries considered are according to the availability of publicly available data in each case.

\subsection{Distribution of Population for different countries }

\begin{figure}[h]
	\subfloat[Estimates of parameters $(a,b)$]{
		\includegraphics[width=0.5\textwidth]{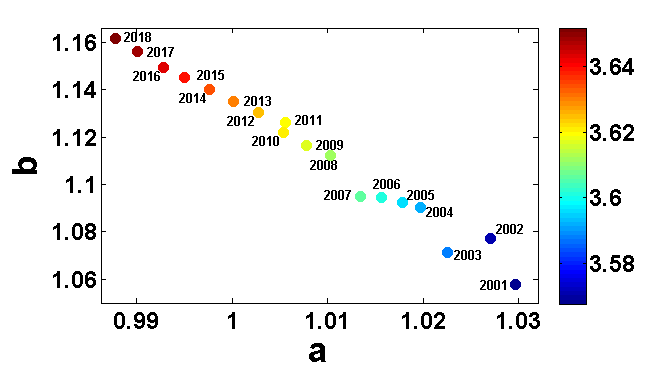}
		\label{FIG:PopWorld_ab}
	}
	\subfloat[Estimated Entropy $\widehat{S}_{\max}$]{
		\includegraphics[width=0.5\textwidth]{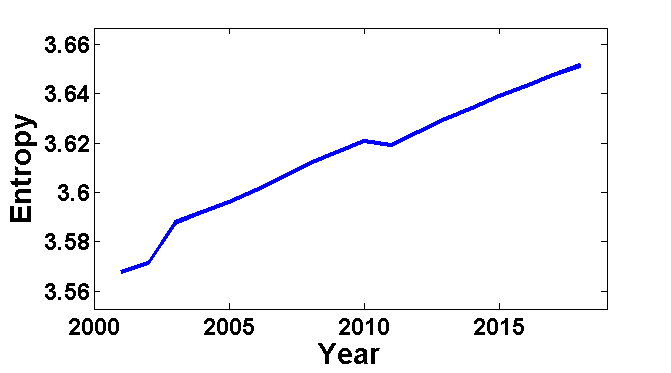}
		\label{FIG:PopWorld_ent}
	}
\caption{Estimates of the parameters $(a,b)$ and the maximum entropy values $\widehat{S}_{\max}$  for the MaxEnt RO distribution 
	fitted  to the countries' population data  over different years} 
	\label{FIG:PopWorld}
\end{figure}

As the first illustration of our global modelling, we have studied the distribution of the country sizes in terms of their population
for each year from 2001 to 2018 obtained from \cite{pop}; these cover around 190 major countries and/or territories   across the world.
The excellent fit of the MaxEnt RO distribution for this case can again be noted  from Supplementary Table \ref{TAB: POP} and Figure 13;
the KS error ranges from 0.0558 to 0.0588 only. The parameter estimates are plotted in Figure \ref{FIG:PopWorld} along with the corresponding entropy.
It is interesting that the countries around the world have a similar distributional structure 
 as that of the  cities within a country and we should be able to explain them as well with a city size law; see also \cite{physica19}. 
However, unlike Japan's city-size distribution, the entropy of the country-size distribution is increasing over time 
indicating the decreasing discrimination between the population in different countries. 
The change is, although, significantly slower with values ranging from 3.5678 to 3.6516 only over the last 18 years
and still significantly away from the uniformity (corresponding global maximum of entropy is $\log(N)$ = 5.247).
The shape of the distribution is also changing gradually  in a fixed pattern with decreasing $a$ and increasing $b$.
Additionally, the changes in countries' population distribution are observed to be significant around the years 2003 and 2011, 
which would be a interesting topic for further investigation.

\subsection{Distribution of Agricultural land across countries  }

We now discuss a fascinating  socio-economic indicator, probably not studied previously,  
namely the percentage of available agricultural land in different countries across the world.  This study may help the Policymakers to create a better and more coherent policy environment to meet food demand sustainably.
Agricultural land, defined by the Food and Agriculture Organization (FAO), refers to the share of land area that is arable, 
under permanent crops, and under permanent pastures. 
%
We have considered the data from \cite{agri}, collected annually by FAO, 
over a time window of 14 years (2003 -- 2016)  for around 192-194  countries as per the availability. 

\begin{figure}[h]
	\subfloat[Estimates of parameters $(a,b)$]{
		\includegraphics[width=0.5\textwidth]{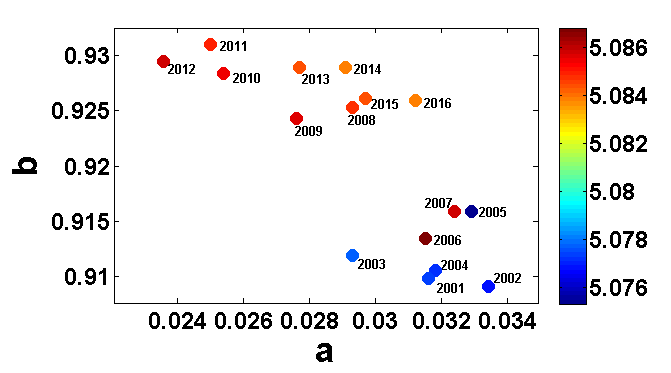}
		\label{FIG:Agriland_ab}
	}
	\subfloat[Estimated Entropy $\widehat{S}_{\max}$]{
		\includegraphics[width=0.4\textwidth]{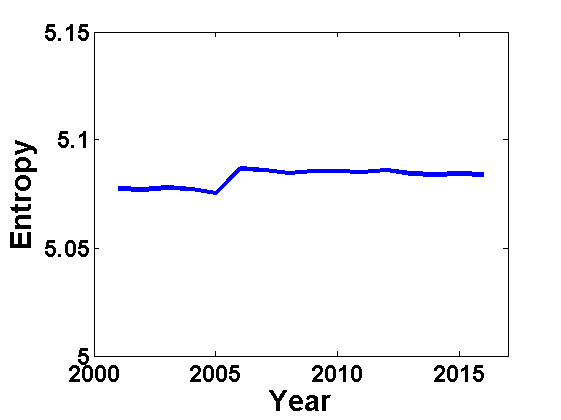}
		\label{FIG:Agriland_ent}
	}
\caption{Estimates of the parameters $(a,b)$ and the maximum entropy values $\widehat{S}_{\max}$  
	for the MaxEnt RO distribution fitted  to the Agricultural land percentage data  over different years} 
	\label{FIG:Agriland}
\end{figure}

Once again we apply the proposed MaxEnt approach with the DGB distribution which is seen to fit the empirical data exceptionally well with a very small KS error of 0.0044 -- 0.0074;
see Supplementary Table \ref{TAB: AGRILAND} and Figure 14 for detailed results and model fits for each year.
Remarkably here, the entropy value $\widehat{S}_{\max}$ remains mostly stable over the years 
(except for sudden increase around the year 2006-07); it varies over 5.05 -- 5.09 only.
However, the distributional shapes, studied through the estimated values of $(a,b)$ in Figure \ref{FIG:Agriland},
are different in most years and form  two prominent clusters before and after the years 2007. 
It would be interesting to investigate such a distributional change in world's agricultural land percentage from 2007!
We may point out here that  worldwide food prices increased dramatically in 2007 and the first and second quarter of 2008 \cite{wfs}, creating a global crisis and causing political and economic instability and social unrest in both poor and developed nations. 

\subsection{Per capita GDP for various countries over different years }

Another very important global socio-economic indicator is the Gross domestic product (GDP), 
a national accounts indicator of a country's economic performance and strength. 
It is measured by the added value of all final goods and services produced in a country 
during a specific time period or by adding every person’s income during that time period. 
Per capita GDP is also used to describe the standard of living of a population, 
with a higher GDP meaning a higher standard of living. 
Here, we have used the publicly available data on the per capita GDP of about 100 countries
for the years 2001--2018, according to their availability from \cite{gdp}, 
and applied the proposed MaxEnt approach. As can be seen from the detailed results given in Supplementary Table \ref{TAB: GDP}
and Figure 15, the DGB distribution again yields very good fit for the per capita GDP data across the world
in each year -- the KS ranges in [0.0136--0.0293] only.



\begin{figure}[h]
	\subfloat[Estimates of parameters $(a,b)$]{
		\includegraphics[width=0.5\textwidth]{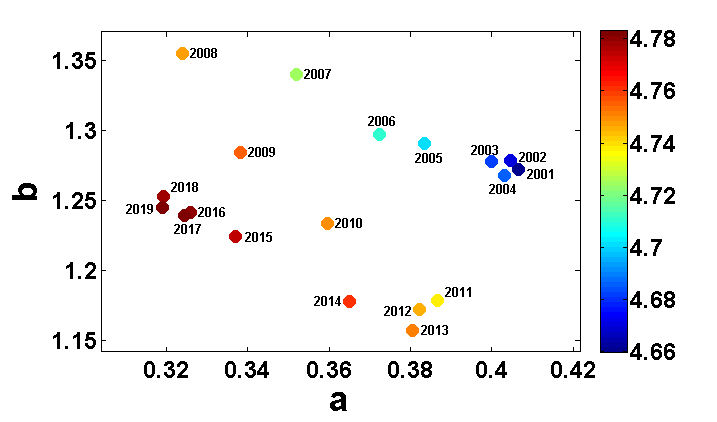}
		\label{FIG:GDP_ab}
	}
	\subfloat[Estimated Entropy $\widehat{S}_{\max}$]{
		\includegraphics[width=0.5\textwidth]{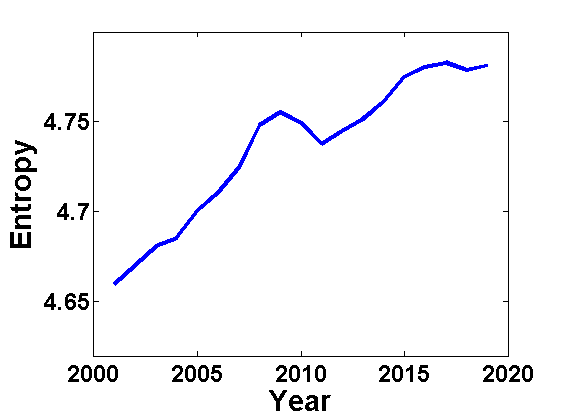}
		\label{FIG:GDP_ent}
	}
\caption{Estimates of the parameters $(a,b)$ and the maximum entropy values $\widehat{S}_{\max}$  for the MaxEnt RO distribution 	fitted  to the per capita GDP data  over different years} 
	\label{FIG:GDP}
\end{figure}

From the plots of the estimated parameters $(a,b)$ and the corresponding maximum entropy $\widehat{S}_{\max}$
in Figure \ref{FIG:GDP}, a clear trend of the entropy  is visible over the years 
although no pattern emerges for distributional shapes. 
On the whole, from the years 2001 to 2018, world's per capita GPD (country-wise) seems to have an increasing trend
of the underlying entropy measure indicating the decreasing inequality among countries' economic strengths.
However, an interesting phenomenon is also observed around the year 2010
when the entropy abnormally break the pattern to decrease from 2009 to 2011. Without a detailed analysis, we are not in a position to remark whether this result is due to the 2007-2010 US financial crisis.

\subsection{Buffet Indicators 
of different countries across the world  }

Our last illustration is to emphasis that the proposed MaxEnt approach is 
not necessarily limited to count, continuous or percentage data only;
rather it can also be used to successfully model the ratio-type socio-economic indicators as well. 
Here, for example, we consider the Buffet indicators of different countries 
that describe the ratio of Market Capitalization  and \% of GDP.
This financial market indicator signifies the percentage of GDP that represents stock market value
and used to determine whether an overall market is undervalued or overvalued compared to a historical average. 
This is probably the best single measure of where valuations stand at any given moment. It is a measure of the total value of all publicly traded stocks in a market divided by that economy's gross domestic product (GDP). The ratio compares the value of all stocks at an aggregate level to the value of the country's total output. 

\begin{figure}[h]
	\subfloat[Estimates of parameters $(a,b)$]{
		\includegraphics[width=0.5\textwidth]{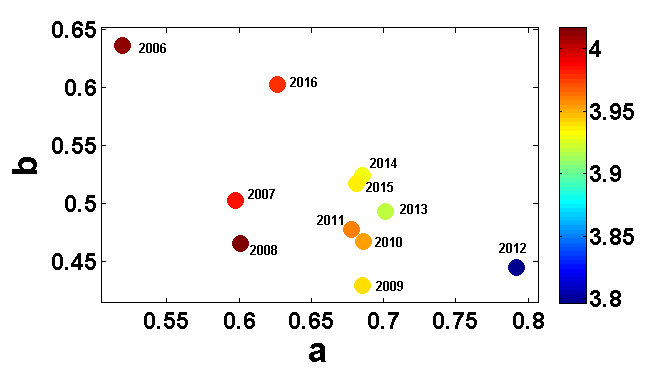}
		\label{FIG:Buffetind_ab}
	}
	\subfloat[Estimated Entropy $\widehat{S}_{\max}$]{
		\includegraphics[width=0.5\textwidth]{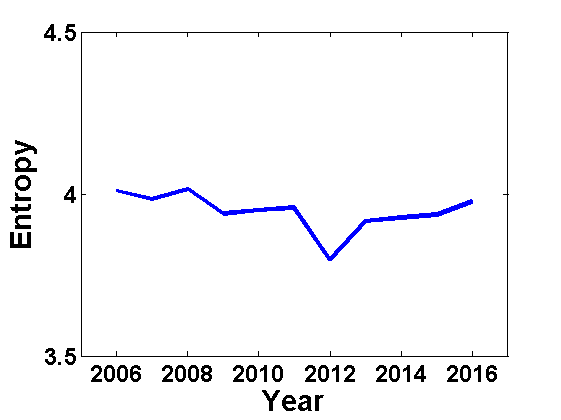}
		\label{FIG:Buffetind_ent}
	}
\caption{Estimates of the parameters $(a,b)$ and the maximum entropy values $\widehat{S}_{\max}$  for the MaxEnt RO distribution 
	fitted  to the  data of Buffet Indicators of various countries over different years} 
	\label{FIG:Buffetind}
\end{figure}

Here, we have used the data for around 190 countries from the year 2006 to 2016, as available in  \cite{market}.
Results of the MaxEnt analyses of these data are given in Supplementary Table \ref{TAB:Buffet} and Figure 16;
the KS values  ranges in [0.0385, 0.0621] indicating yet another example of excellent DGB fit.
However, unlike the previous cases, there is no prominent pattern in the fitted RO distributions 
to the Buffet indicators over different years both in terms of parameter estimates and the entropy values; 
see Figure \ref{FIG:Buffetind}. 
We should note here a potential issue with the Buffett Indicator; 
the underlying assumption of stable corporate earnings relative to economic activity may be wrong, 
or it may be correct for the United States but not for other markets. 
More investigation is required for further in-depth analyses. 

\section{Discussions}
Recently, there was a proposal showing a universal behavior defined in terms of a functional relation for rank ordered distributions that holds accurately along the whole rank range for an impressive amount of phenomena of very diverse nature \cite{Martinez-Mekler09}. This distribution goes beyond power laws since it is a two parameter relation that reduces to a power law when one of them is zero. It is observed that this distribution is very much useful in modelling many socio-economic factors providing  excellent fits when validated with real world empirical data. However, the underlying theory responsible for the success of this particular RO distribution is not explored properly. This paper is the first to provide a derivation of this particularly useful RO distribution
as a natural $maximum$ $entropy$ $distribution$  under the  appropriate bivariate utility constraints. 
Moreover, here  we have also illustrated its acceptability in universal modeling of different types of socio-economic factors. 
In particular, we have considered data for diverse socio-economic  factors within a country as well as across the countries to demonstrate our theory. 
In this maximum entropy framework, the  two parameters corresponding to the distribution of various socio-economic indicators can be thought of as 
two $sensors$ or $ latent drivers $  of the respective dynamics and  may further be associated with observable economic, social or environmental factors.
The  estimated values of distributional parameters, $sensors$,  along with the $ entropy$ 
helped us to investigate  the underlying theory of the distribution. 
The evolution of the $entropy$ over a time span showed how the underlying uncertainty in the distribution changes over time. 
It is very fascinating to note that the world food crisis in 2007 has an inherent imprint over the percentage of agricultural land data analysis (See Fig. 5). 
Similarly, the entropy analysis of GDP per capita and Buffet indicator of various countries over many years (Fig. 6 and 7)  give hints of the world financial crisis during 2007-2010.     
We believe that the universal rank-order modeling, along with the underlying entropy analysis   for diverse socio-economic measures  
around the world will shed light on setting policies  promoting inclusive, equitable and sustainable development.

 \bigskip
 \noindent
 {\bf{Acknowledgment}}:\\
PS acknowledges Indian Statistical Institute, Kolkata, India for allowing her to work as a summer/project  trainee in the Physics and Applied Mathematics Unit.   

\appendix
\section{Method: Fitting the Max-Ent RO distribution to Empirical Data}
\label{APP:A}

Suppose there are $N$ items (data-points) whose ranks and sizes are, respectively, denoted as $r_i$ and $x_i$
for $i=1, \ldots, N$. We need to find a best fitted DGB distribution through 
the estimation of the parameters $(a,b)$ from the empirical data.
We use the maximum likelihood approach to estimate $(a,b)$ by maximizing the likelihood function 
\begin{equation}
L(a, b) = \prod_{i=1}^N f_{(a,b)}(r_i)^{{x}_i} = \prod_{i=1}^N \frac{(N+1-r_i)^{b{x}_i}}{r_i^{a{x}_i}}A^{{x}_i},
\end{equation}
or, equivalently the log-likelihood function $\ell(a, b)  = \log L(a, b)$.
Our numerical illustrations are performed in the software MATLAB (version14a)
using the in-built function \textit{'fminsearch'}; see \cite{physica19} for more details and justifications.

After estimating the parameters as $(\widehat{a}, \widehat{b})$, we compute the predicted size of $r_i$ as 
$p_i = \left(\sum_{i=1}^N x_i\right)f_{(\widehat{a}, \widehat{b})}(r_i),$
and compare it with the observed sizes $x_i$ to investigate the goodness-of-fit of our RO model.
The overall error in this fit is quantified through the Kolmogorov-Smirnov (KS) measure between 
the predicted and the observed cumulative frequencies of ranks:
\begin{eqnarray}
KS = \max_{1\leq i \leq N} \left|\left(\sum_{j: r_j\leq r_i} p_j\right) - \left(\sum_{j: r_j\leq r_i} x_j\right)\right|. 
\label{EQ:KS}
\end{eqnarray}


\newpage
%

\section{Supplementary Tables}

\vspace{2in}

\begin{table}[h]
	\caption{Estimated measures for Japanese  Cities (human settlements) 
		along with total number of cities ($N$)
		and the sizes $x_{\min}, x_{\max}$ of the smallest and the largest cities, respectively.}
	\centering	
	\begin{tabular}{l|rrr|rrr|rr}\hline
		& \multicolumn{3}{c}{Summary}	& \multicolumn{3}{|c|}{RO Fit}  & \multicolumn{1}{c}{Entropy}  \\
		Year	&	$N$	& $x_{\min}$ & $x_{\max}$ &	$\widehat{a}$	&	$\widehat{b}$	&	 $KS$	&	$S$ 	\\ \hline
		2015    &   976 & 16742 & 9272740   &   0.7656  &   0.2487  &   0.0149  &  5.9292 \\
		2010	&	985	& 16628 & 8945695 	&	0.7582	&	0.2399	&	0.0140  &  5.9653 \\
		2005	&	985	& 18060 & 8489653 	&	0.7503	&	0.2283	&	0.0136	&  5.9962 \\
		2000	&	985	& 16504 & 8134688 	&	0.7436	&	0.2225	&	0.0133	&  6.0197 \\
		1995    &   985 & 15171 & 7967614   &   0.7413  &   0.2182  &   0.0134  &  6.0289 \\
		\hline
	\end{tabular}
	\label{TAB:Japan}
\end{table}

\begin{table}[h]
	\caption{Parameter estimates for the distribution of Personal Income for USA, along with total number of income groups ($N$) and the sizes $x_{\min},x_{\max}$ of the smallest and largest income groups respectively.} 
	\centering
	\begin{tabular}{l|rrr|rrr|r}\hline
		&\multicolumn{3}{c}{summary} & \multicolumn{3}{|c|}{RO Fit} & \multicolumn{1}{c}{Entropy} \\
		Year  & $N$  &  $x_{\min}$ &  $x_{\max}$  &  $\widehat{a}$  &  $\widehat{b}$  &  $KS$  & $S$\\ \hline
		2017  &  41  &  0780.0000  &  24926.0  &  0.3751  &  0.6772  &  0.0149  &  3.4338 \\
		2016  &  41  &  0770.1000  &  22426.1  &  0.3381  &  0.7619  &  0.1640  &  3.4310 \\
		2015  &  41  &  0712.0000  &  20755.0  &  0.3053  &  0.8480  &  0.0179  &  3.4250 \\
		2014  &  41  &  0683.0000  &  19063.0  &  0.2811  &  0.9209  &  0.0189  &  3.4178 \\
		2013  &  41  &  0672.0000  &  19108.0  &  0.2825  &  0.9310  &  0.0184  &  3.4138 \\
		2012  &  41  &  0575.8441  &  15727.4  &  0.1975  &  1.0897  &  0.0237  &  3.4179 \\
		\hline
	\end{tabular}
	\label{TAB: USAP}      
\end{table}


\begin{table}[h]
	\caption{Parameter estimates for the distribution of  Household Income for USA, along with total number of income groups ($N$)
		and the sizes $x_{\min}, x_{\max}$ of the smallest and the largest cities, respectively.}
	\centering	
	\begin{tabular}{l|rrr|rrr|r}\hline
		& \multicolumn{3}{c}{Summary}	& \multicolumn{3}{|c|}{RO Fit} & \multicolumn{1}{c}{Entropy} \\
		Year	&	$N$	& $x_{\min}$ & $x_{\max}$ &	$\widehat{a}$	&	$\widehat{b}$	&	$KS$ 	&	$S$	\\ \hline
		2017  &  41  &  588  &  9874  &  0.2344  &  0.7325  &  0.0188  &  3.5021 \\
		2016  &  41  &  482  &  8775  &  0.1874  &  0.8418  &  0.0193	  &  3.4944 \\
		2015  &  41  &  431  &  7640  &  0.1595  &  0.9233  &  0.0188	  &  3.4851 \\
		2014	  &  41  &  436  &  7005  &  0.1404  &  1.0173  &  0.0243	  &  3.4683 \\
		2013  &  41  &  347  &  7085  &  0.1358  &  1.0536  &  0.0255  &  3.4606 \\
		2012  &  41  &  364  &  7157  &  0.1235  &  1.1478  &  0.0243  &  3.4411 \\
		\hline
	\end{tabular}
	\label{TAB: USAH}
\end{table}

   \begin{table}[h]
	\caption{Parameter estimates for the distribution of vote percentage share of winner candidate in terms of percentage for India... in their respective constituency along with total number of Parliamentary constituencies ($N$)and
		the sizes $x_{\min},x_{\max}$ of the smallest and largest percentage of winners' vote percentage respectively.}
	\centering
	\begin{tabular}{l|rrr|rrr|r}\hline
		& \multicolumn{3}{c}{Summary}  &  \multicolumn{3}{|c|}{RO Fit}  &  \multicolumn{1}{c}{Entropy} \\
		Year  &  $N$  &  $x_{\min}$  &  $x_{\max}$  & $\widehat{a}$  &  $\widehat{b}$  &  $KS$  &  $S$\\ \hline
		
		2019  &  534  &  32.40\%  &  74.47\%  &  0.0664  &  0.0975  &  0.0021  &  6.2701  \\
		2014  &  535  &  27.30\%  &  75.79\%  &  0.0840  &  0.1132  &  0.0035  &  6.2673  \\
		2009  &  535  &  14.19\%  &  78.80\%  &  0.0754  &  0.1521  &  0.0019  &  6.2631  \\
		2004  &  398  &  25.98\%  &  77.16\%  &  0.0747  &  0.1544  &  0.0036  &  5.9672  \\
		1999  &  371  &  20.83\%  &  74.91\%  &  0.0630  &  0.1583  &  0.0027  &  5.8984  \\
		1998  &  371  &  25.55\%  &  86.70\%  &  0.0813  &  0.1058  &  0.0028  &  5.9028  \\
		1996  &  371  &  24.36\%  &  78.47\%  &  0.0838  &  0.1452  &  0.0037  &  5.8967  \\
		1991  &  363  &  23.06\%  &  90.12\%  &  0.0932  &  0.1667  &  0.0030  &  5.8697  \\
		1989  &  351  &  27.15\%  &  84.08\%  &  0.0678  &  0.1463  &  0.0024  &  5.8440  \\
		1984  &  363  &  25.81\%  &  83.67\%  &  0.0687  &  0.1192  &  0.0027  &  5.8812  \\
		1980  &  351  &  22.68\%  &  77.53\%  &  0.0717  &  0.1460  &  0.0032  &  5.8434  \\ 	
		\hline
	\end{tabular}

	\label{TAB:ECIWP}
\end{table}


\begin{table}[h]
	\caption{Parameter estimates for the distribution of vote percentage share margin of the winner candidate in terms of percentage for India in their respective constituency along with total number of Parliamentary Constituencies ($N$)
		and the sizes $x_{\min}, x_{\max}$ of the smallest and largest percentage of winners' vote margin percentage respectively.}
	\centering
	\begin{tabular}{l|rrr|rrr|r}\hline
		& \multicolumn{3}{c}{Summary}  &  \multicolumn{3}{|c|}{RO Fit}  &  \multicolumn{1}{c}{Entropy} \\
		Year  &  $N$  &  $x_{\min}$  &  $x_{\max}$  &  $\widehat{a}$  &  $\widehat{b}$  &  $KS$  &  $S$ \\ \hline		
		2019  &  534  &  0.02\%  &  52.73\%  &  0.0993  &  1.0916  &  0.0029  &  6.0176  \\
		2014  &  535  &  0.07\%  &  56.25\%  &  0.1435  &  0.9190  &  0.0056  &  6.0422  \\
		2009  &  535  &  0.04\%  &  70.10\%  &  0.3019  &  0.9349  &  0.0055  &  5.9162  \\
		2004  &  398  &  0.06\%  &  61.41\%  &  0.2480  &  1.0443  &  0.0054  &  5.6363  \\
		1999  &  371  &  0.13\%  &  53.54\%  &  0.2465  &  0.9926  &  0.0031  &  5.5839  \\
		1998  &  371  &  0.01\%  &  73.41\%  &  0.2758  &  0.9999  &  0.0055  &  5.5568  \\
		1996  &  371  &  0.06\%  &  72.00\%  &  0.2363  &  0.9947  &  0.1170  &  5.5914  \\
		1991  &  363  &  0.03\%  &  87.19\%  &  0.2595  &  1.0202  &  0.0072  &  5.5429  \\
		1989  &  349  &  0.16\%  &  70.30\%  &  0.1828  &  0.8870  &  0.0034  &  5.6016  \\
		1984  &  362  &  0.02\%  &  72.18\%  &  0.1195  &  1.0437  &  0.0084  &  5.6315  \\
		1980  &  351  &  0.02\%  &  68.49\%  &  0.1316  &  1.0053  &  0.0036  &  5.6045  \\  		
		\hline
	\end{tabular}
	\label{TAB:ECIWMP}
\end{table}

\vspace{-.1in}
\begin{table}[h]
	\caption{Parameter Estimates for Buffet indicator, along with number of countries ($N$)
		and the entries $x_{\min}, x_{\max}$ }
	\centering	
	\begin{tabular}{l|rrr|rrr|r}\hline
		& \multicolumn{3}{c}{Summary}	& \multicolumn{3}{|c|}{RO Fit} & \multicolumn{1}{c}{Entropy} \\
		Year	&	$N$	& $x_{\min}$ & $x_{\max}$ &	$\widehat{a}$	&	$\widehat{b}$	&	KS	&	S \\ \hline
		2016 & 93 & 0.083 & 0981.639 & 0.6267  & 0.6023	 & 0.0535 & 3.9772 \\
		2015 & 93 & 0.092 & 1018.500 & 0.6812  & 0.5170  & 0.0494 & 3.9373 \\
		2014 & 93 & 0.086 & 1101.432 & 0.6854  & 0.5236	 & 0.0581 & 3.9282 \\
		2013 & 93 & 0.084 & 1118.119 & 0.7014  & 0.4933  & 0.0592 & 3.9179 \\
		2012 & 93 & 0.101 & 1073.692 & 0.7918  & 0.4443  & 0.0467 & 3.7964 \\
		2011 & 93 & 0.101 & 0902.212 & 0.6776  & 0.4778  & 0.0564 & 3.9596 \\
		2010 & 93 & 0.066 & 1178.962 & 0.6861  & 0.4672  & 0.0621 & 3.9519 \\
		2009 & 90 & 0.066 & 1070.925 & 0.6857  & 0.4292  & 0.0569 & 3.9406 \\
		2008 & 88 & 0.059 & 0600.508 & 0.6012  & 0.4656  & 0.0366 & 4.0175 \\
		2007 & 86 & 0.069 & 1244.064 & 0.5975  & 0.5024  & 0.0545 & 3.9857 \\
		2006 & 85 & 0.079 & 0881.286 & 0.5195  & 0.6360  & 0.0385 & 4.0118 \\
		\hline
	\end{tabular}
	\label{TAB:Buffet}
\end{table}

\clearpage
\begin{table}[h]
	\caption{Estimation of parameters for the distribution of population of various countries in the world, along with number of countries ($N$)
		and the entries $x_{\min}, x_{\max}$ for the population}
	\centering	
	\begin{tabular}{l|rrr|rrr|r}\hline
		& \multicolumn{3}{c}{Summary}	& \multicolumn{3}{|c|}{RO Fit} & \multicolumn{1}{c}{Entropy} \\
		Year	  &	$N $  &  $x_{\min}$(Mn)  &  $x_{\max}$(Mn)  &  $\widehat{a}$  &  $\widehat{b}$  &  $KS$ &  $S$  \\ \hline
		2018  &  192  &  0.011  & 1396.982  &  0.9878  &  1.1614  &  0.0588  &  3.65160  \\ 	
		2017  &  192  &  0.011  & 1390.080  &  0.9901  &  1.1560  &  0.0587  &  3.64770  \\
		2016  &  192  &  0.011  & 1382.710  &  0.9928  &  1.1493  &  0.0585  &  3.64330  \\
		2015  &  192  &  0.011  & 1374.620  &  0.9950  &  1.1452  &  0.0584  &  3.63910  \\
		2014  &  192  &  0.011  & 1367.820  &  0.9976  &  1.1399  &  0.0582  &  3.63440  \\
		2013  &  192  &  0.011  & 1360.720  &  1.0001  &  1.1351  &  0.0581  &  3.62970  \\
		2012  &  192  &  0.010  & 1354.040  &  1.0028  &  1.1303  &  0.0580  &  3.62470  \\
		2011  &  192  &  0.010  & 1347.350  &  1.0056  &  1.1260  &  0.0579  &  3.61910  \\
		2010  &  192  &  0.010  & 1340.910  &  1.0054  &  1.1219  &  0.0581  &  3.62090  \\
		2009  &  192  &  0.010  & 1334.500  &  1.0078  &  1.1166  &  0.0580  &  3.61652  \\
		2008  &  192  &  0.009  & 1328.020  &  1.0103  &  1.1121  &  0.0579  &  3.61180  \\
		2007  &  191  &  0.009  & 1321.290  &  1.0134  &  1.0951  &  0.0577  &  3.60620  \\
		2006  &  191  &  0.009  & 1314.480  &  1.0156  &  1.0944  &  0.0575  &  3.60090  \\
		2005  &  191  &  0.010  & 1307.560  &  1.0178  &  1.0925  &  0.0573  &  3.59620  \\
		2004  &  191  &  0.010  & 1299.880  &  1.0197  &  1.0901  &  0.0571  &  3.59210  \\
		2003  &  190  &  0.009  & 1292.270  &  1.0225  &  1.0713  &  0.0567  &  3.58790  \\
		2002  &  189  &  0.009  & 1284.530  &  1.0270  &  1.0771  &  0.0562  &  3.57140  \\
		2001  &  188  &  0.019  & 1276.270  &  1.0296  &  1.0577  &  0.0558  &  3.56780  \\		
		\hline
	\end{tabular}
	\label{TAB: POP}
\end{table}

\begin{table}[h]
	\caption{Parameter estimates for the distribution of percentage of agricultural land in various countries across the world, along with number of countries ($N$)
		and the entries $x_{\min}, x_{\max}$ maximum and minimum land percentage in that particular year respectively. }
	\centering	
	\begin{tabular}{l|rrr|rrr|r}\hline
		& \multicolumn{3}{c}{Summary}	& \multicolumn{3}{|c|}{RO Fit} & \multicolumn{1}{c}{Entropy} \\
		Year  & $N$  &  $x_{\min}$  &  $x_{\max}$  &	 $\widehat{a}$  &  $\widehat{b}$  &  $KS$  &  $S$  \\ \hline		
		2016  &  194  &  0.56\%  &  82.56\%  &  0.0312  &  0.9259  &  0.0072  &  5.0838  \\
		2015  &  194  &  0.56\%  &  82.56\%  &  0.0297  &  0.9261  &  0.0072  &  5.0843  \\
		2014  &  194  &  0.57\%  &  82.56\%  &  0.0291  &  0.9289  &  0.0072  &  5.0839  \\
		2013  &  194  &  0.53\%  &  82.64\%  &  0.0277  &  0.9289  &  0.0070  &  5.0844  \\
		2012  &  194  &  0.47\%  &  81.30\%  &  0.0236  &  0.9295  &  0.0072  &  5.0859  \\
		2011  &  194  &  0.51\%  &  83.00\%  &  0.0250  &  0.9310  &  0.0074  &  5.0850  \\
		2010  &  194  &  0.50\%  &  82.46\%  &  0.0254  &  0.9284  &  0.0071  &  5.0855  \\
		2009  &  194  &  0.52\%  &  84.64\%  &  0.0276  &  0.9243  &  0.0067  &  5.0856  \\
		2008  &  194  &  0.45\%  &  83.84\%  &  0.0293  &  0.9253  &  0.0068  &  5.0847  \\
		2007  &  194  &  0.45\%  &  83.13\%  &  0.0324  &  0.9159  &  0.0062  &  5.0859  \\
		2006  &  194  &  0.47\%  &  83.96\%  &  0.0315  &  0.9134  &  0.0053  &  5.0868  \\
		2005  &  192  &  0.47\%  &  84.74\%  &  0.0329  &  0.9159  &  0.0044  &  5.0753  \\
		2004  &  192  &  0.50\%  &  84.73\%  &  0.0318  &  0.9105  &  0.0057  &  5.0571  \\
		2003  &  192  &  0.52\%  &  85.29\%  &  0.0293  &  0.9119  &  0.0062  &  5.0778  \\
		2002  &  192  &  0.47\%  &  85.26\%  &  0.0334  &  0.9091  &  0.0063  &  5.0769  \\
		2001  &  192  &  0.55\%  &  85.49\%  &  0.0316  &  0.9098  &  0.0057  &  5.0774  \\
		\hline
	\end{tabular}
	\label{TAB: AGRILAND}
\end{table}
\clearpage

\begin{table}[h]
	\caption{Parameter estimates for GDP per capita, along with number of countries ($N$)
		and the entries $x_{\min}, x_{\max}$ for the GDP per capita, respectively.}
	\centering	
	\begin{tabular}{l|rrr|rrr|r}\hline
		& \multicolumn{3}{c}{Summary}	& \multicolumn{3}{|c|}{RO Fit} & \multicolumn{1}{c}{Entropy} \\
		Year  &  $N$  &  $x_{\min}$  &  $x_{\max}$  &  $\widehat{a}$  &  $\widehat{b}$  &  $KS$  &  $S$  \\ \hline		
	  2019  &  190  &  726.885  &  134622.6  &  0.3190  &  1.2451  &  0.0176  &  4.7813  \\									
2018  &  190  &  711.896  &  130475.1  &  0.3193  &  1.2527  &  0.0175  &  4.7787  \\
2017  &  191  &  680.323  &  127785.4  &  0.3244  &  1.2394  &  0.0172  &  4.7830  \\
2016  &  191  &  652.481  &  125307.6  &  0.3260  &  1.2416  &  0.0166  &  4.7807  \\
2015  &  191  &  629.558  &  130319.6  &  0.3370  &  1.2244  &  0.0165  &  4.7752  \\
2014  &  191  &  606.024  &  136829.7  &  0.3651  &  1.1774  &  0.0142  &  4.7613  \\
2013  &  191  &  600.193  &  142845.6  &  0.3806  &  1.1567  &  0.0122  &  4.7515  \\
2012  &  191  &  593.842  &  146981.8  &  0.3822  &  1.1721  &  0.0126  &  4.7448  \\
2011  &  191  &  560.425  &  145723.8  &  0.3866  &  1.1786  &  0.0149  &  4.7379  \\
2010  &  191  &  530.043  &  127195.7  &  0.3595  &  1.2338  &  0.0191  &  4.7494  \\
2009  &  191  &  504.838  &  111399.6  &  0.3382  &  1.2844  &  0.0225  &  4.7555  \\
2008  &  191  &  502.787  &  104145.4  &  0.3239  &  1.3556  &  0.0244  &  4.7482  \\
2007  &  191  &  479.299  &  117053.0  &  0.3520  &  1.3403  &  0.0245  &  4.7243  \\
2006  &  190  &  453.592  &  115071.8  &  0.3725  &  1.2970  &  0.0253  &  4.7107  \\
2005  &  190  &  431.732  &  104311.6  &  0.3834  &  1.2910  &  0.0272  &  4.7006  \\
2004  &  190  &  420.928  &  107274.6  &  0.4031  &  1.2681  &  0.0270  &  4.6856  \\
2003  &  189  &  409.659  &  95604.29  &  0.4000  &  1.2778  &  0.0284  &  4.6810  \\
2002  &  188  &  406.197  &  94876.81  &  0.4045  &  1.2788  &  0.0295  &  4.6705  \\
2001  &  186  &  413.190  &  89695.37  &  0.4066  &  1.2722  &  0.0293  &  4.6598  \\
		\hline
	\end{tabular}
	\label{TAB: GDP}
\end{table}

\section{Supplementary Figures}


\begin{figure}[h]
\vspace{-.4in}
	\subfloat[Year 2015]{
	\includegraphics[width=0.3\textwidth]{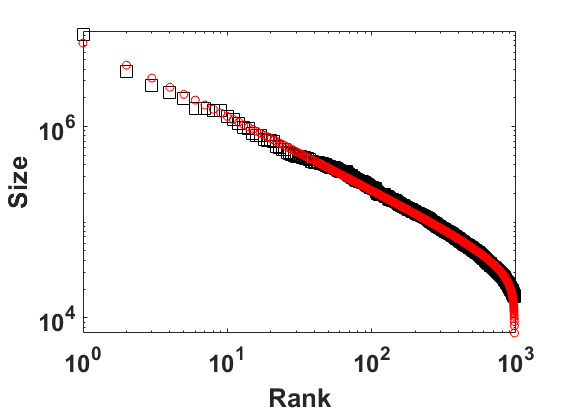}}
	\subfloat[Year 2010]{
	\includegraphics[width=0.3\textwidth]{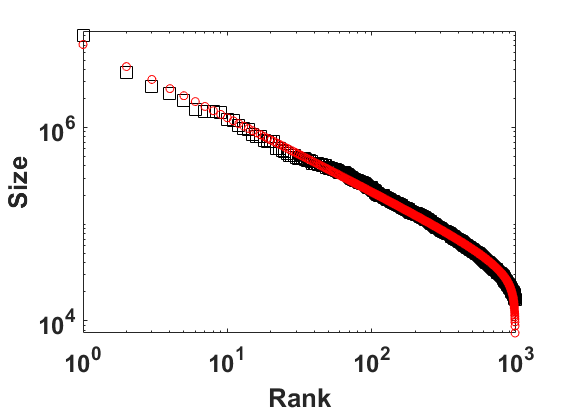}}
	\subfloat[Year 2005]{
	\includegraphics[width=0.3\textwidth]{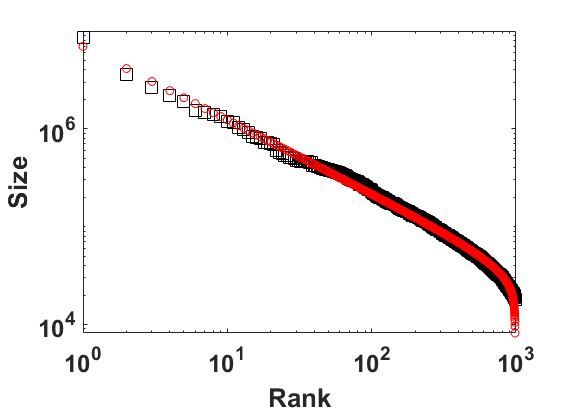}}
\\
	\subfloat[Year 2000]{
	\includegraphics[width=0.3\textwidth]{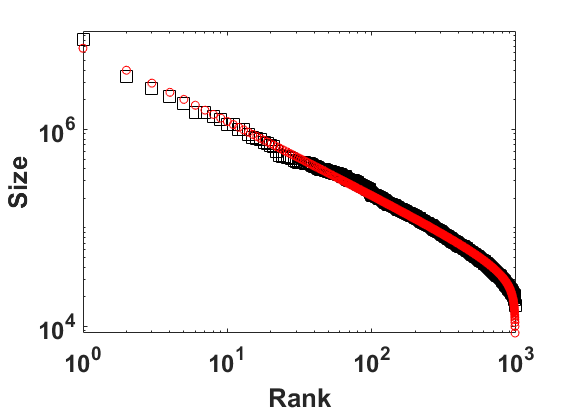}}
	\subfloat[Year 1995]{
	\includegraphics[width=0.3\textwidth]{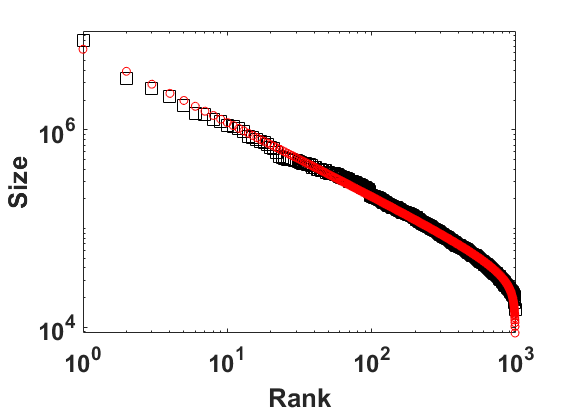}}
	\label{japan2015}
	\caption{Plots of actual (Black square) and predicted (Red circle) sizes over the rank 
		for the city size of Japan} 
\end{figure}

\begin{figure}[h]
	\subfloat[Year 2017]{
	\includegraphics[width=0.3\textwidth]{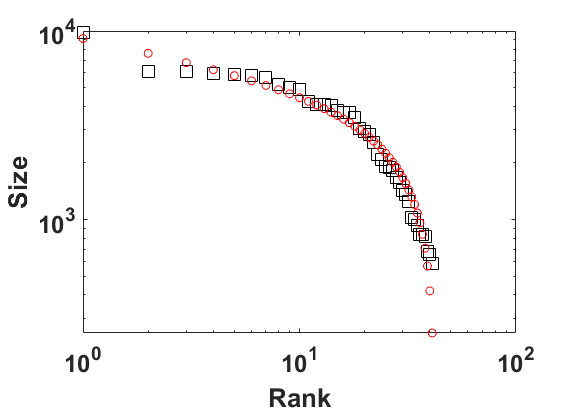}}
	\subfloat[Year 2016]{
	\includegraphics[width=0.3\textwidth]{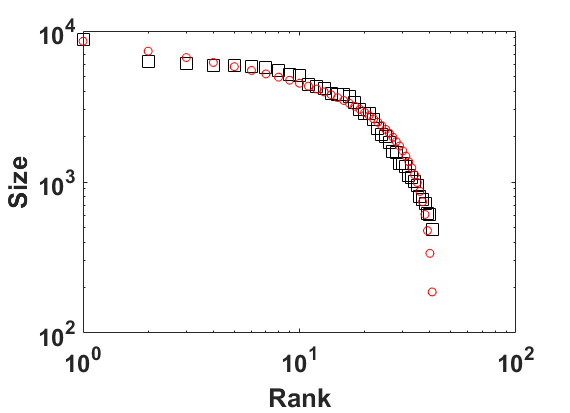}}
	\subfloat[Year 2015]{
	\includegraphics[width=0.3\textwidth]{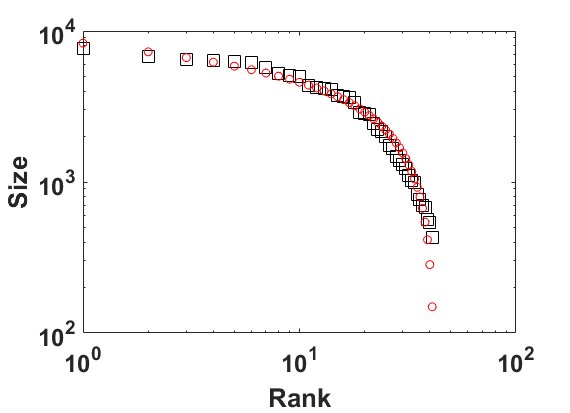}}
\\
	\subfloat[Year 2014]{
	\includegraphics[width=0.3\textwidth]{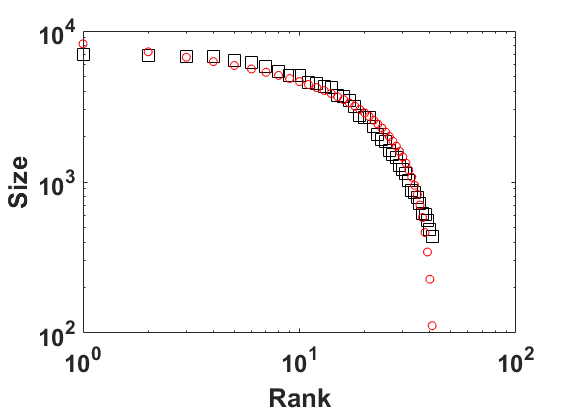}}
	\subfloat[Year 2013]{
	\includegraphics[width=0.3\textwidth]{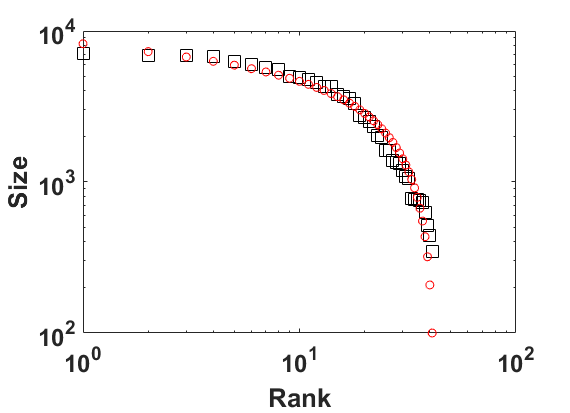}}
	\subfloat[Year 2012]{
	\includegraphics[width=0.3\textwidth]{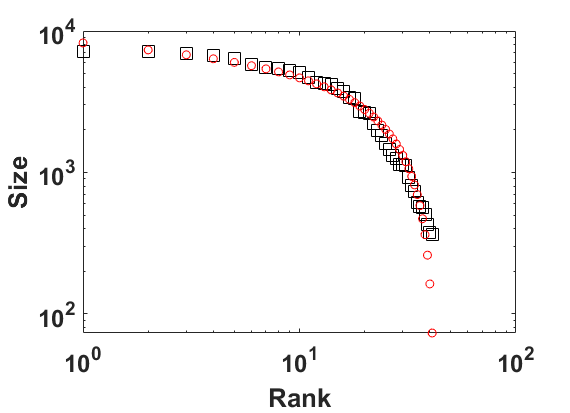}}
	\label{USA_house}
\caption{Plots of actual (Black square) and predicted (Red circle) sizes over the rank 
	for the household income of USA} 
\end{figure}

\begin{figure}[h]
	\subfloat[Year 2017]{
	\includegraphics[width=0.3\textwidth]{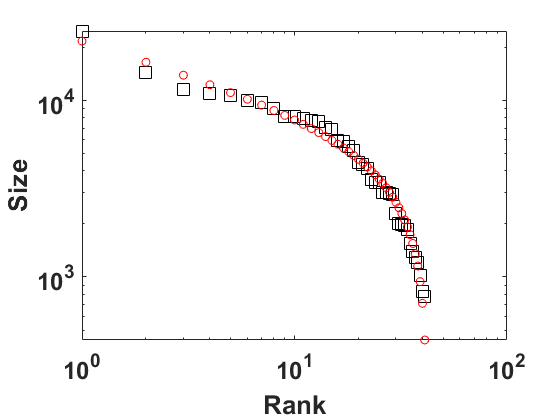}}
	\subfloat[Year 2016]{
	\includegraphics[width=0.3\textwidth]{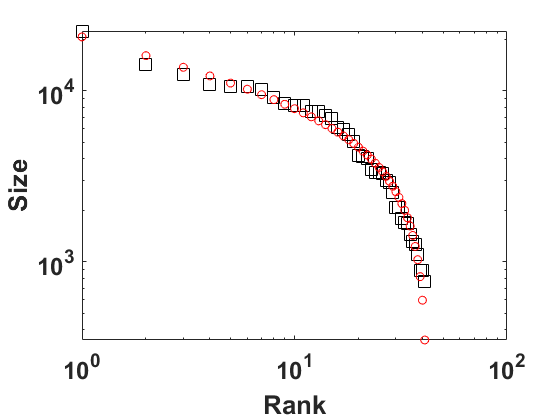}}
	\subfloat[Year 2015]{
	\includegraphics[width=0.3\textwidth]{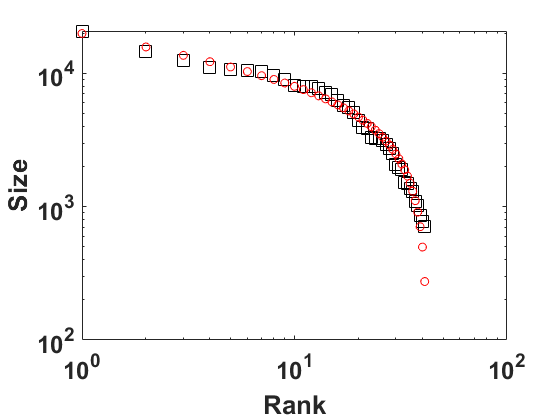}}
\\
	\subfloat[Year 2014]{
	\includegraphics[width=0.3\textwidth]{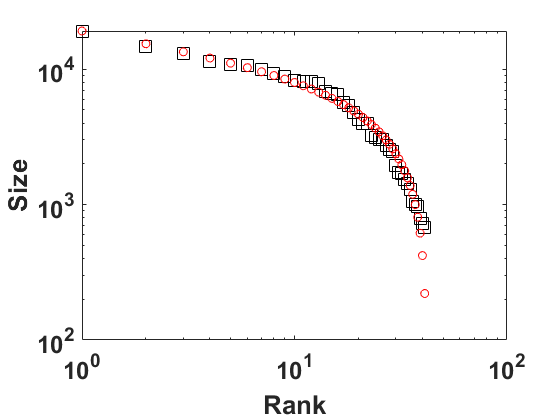}}
	\subfloat[Year 2013]{
	\includegraphics[width=0.3\textwidth]{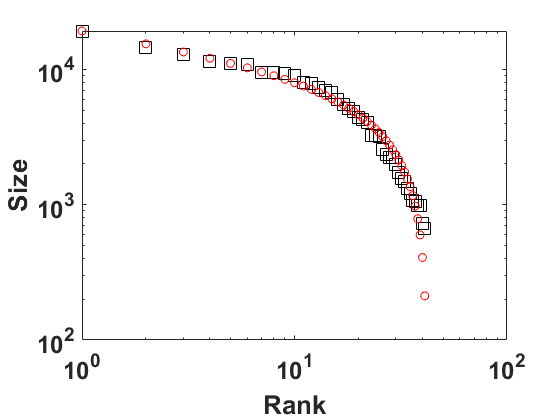}}
	\subfloat[Year 2012]{
	\includegraphics[width=0.3\textwidth]{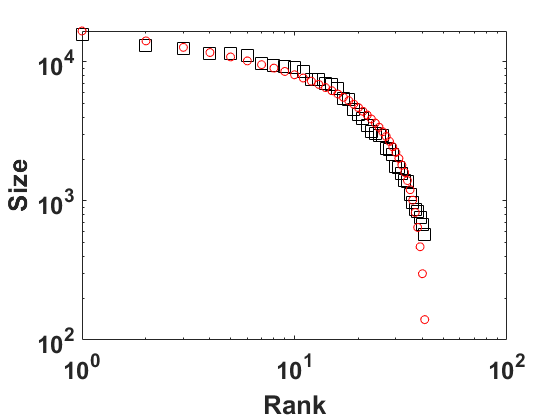}}
	\label{USA_personal}
\caption{Plots of actual (Black square) and predicted (Red circle) sizes over the rank 
	for the personal income of USA} 
\end{figure}

\begin{figure}
	\subfloat[Year 2019]{
	\includegraphics[width=0.3\textwidth]{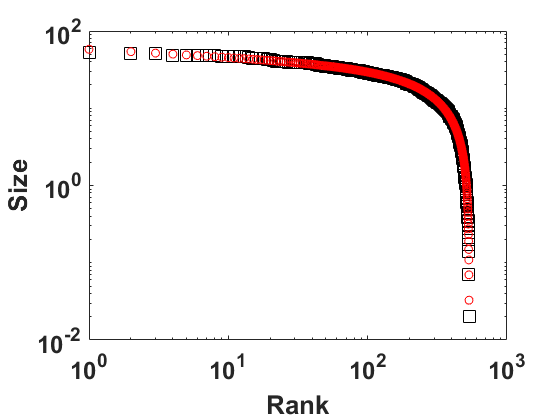}}
	\subfloat[Year 2014]{
	\includegraphics[width=0.3\textwidth]{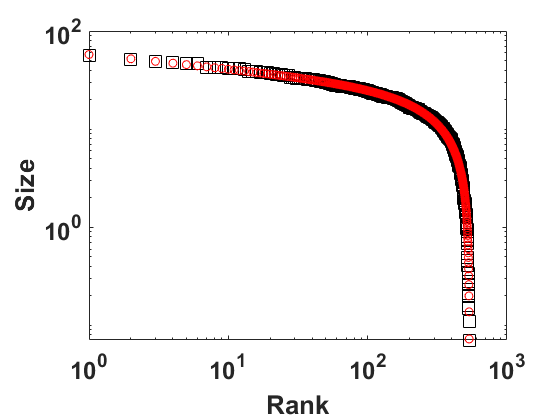}}
	\subfloat[Year 2009]{
	\includegraphics[width=0.3\textwidth]{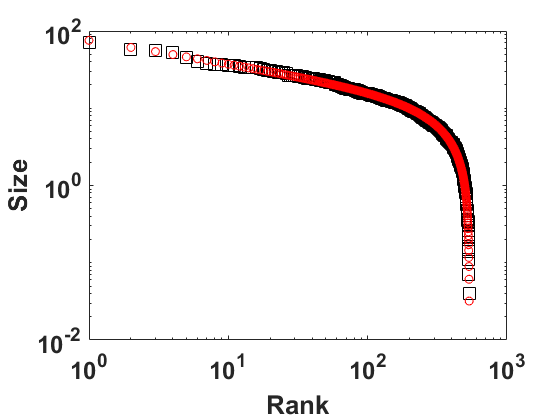}}
\\
	\subfloat[Year 2004]{
	\includegraphics[width=0.3\textwidth]{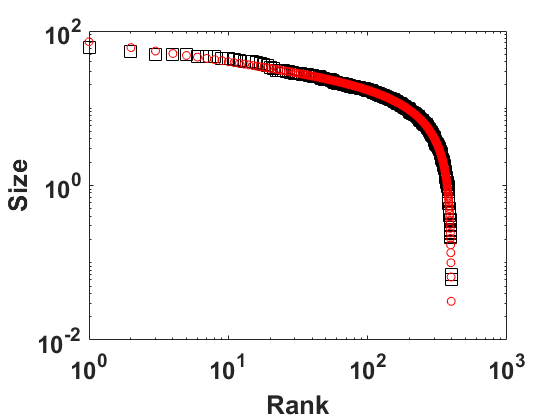}}
	\subfloat[Year 1999]{
	\includegraphics[width=0.3\textwidth]{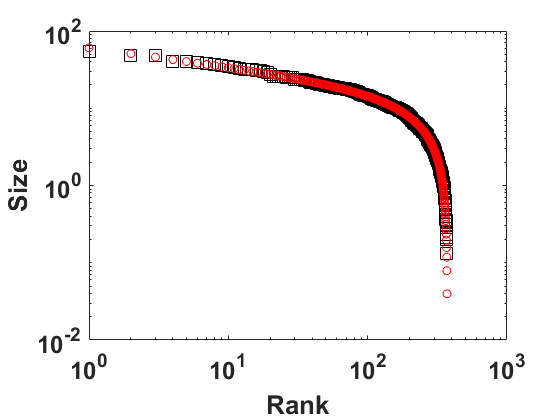}}
	\subfloat[Year 1998]{
	\includegraphics[width=0.3\textwidth]{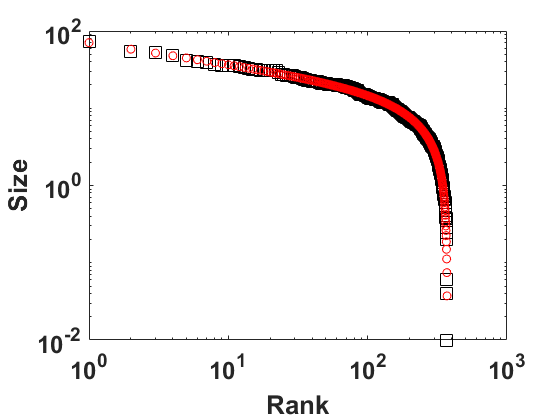}}
\\
	\subfloat[Year 1996]{
	\includegraphics[width=0.3\textwidth]{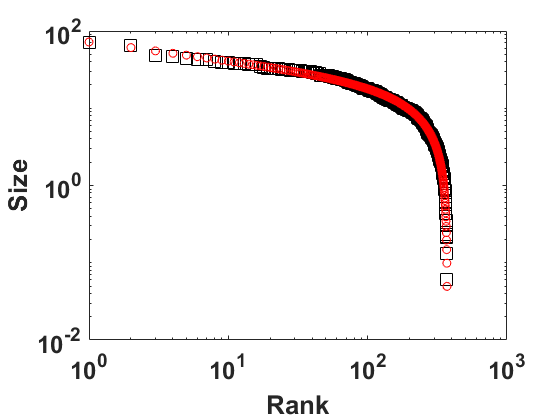}}
	\subfloat[Year 1991]{
	\includegraphics[width=0.3\textwidth]{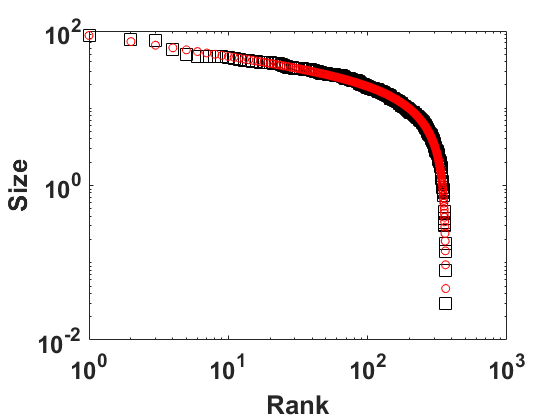}}
	\subfloat[Year 1989]{
	\includegraphics[width=0.3\textwidth]{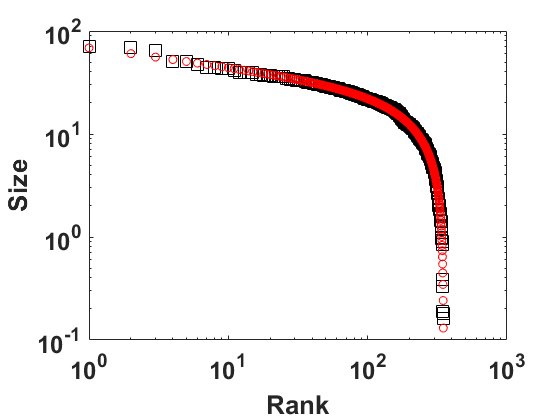}}
\\
	\subfloat[Year 1984]{
	\includegraphics[width=0.3\textwidth]{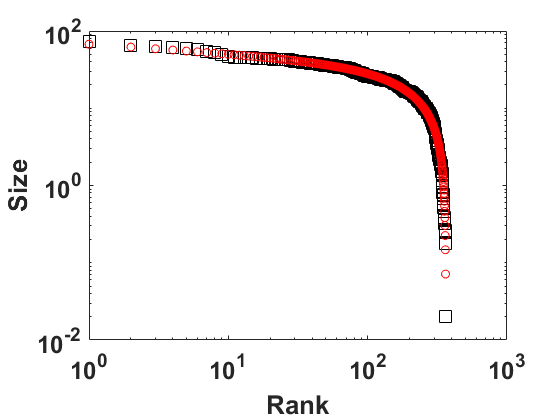}}
	\subfloat[Year 1980]{
	\includegraphics[width=0.3\textwidth]{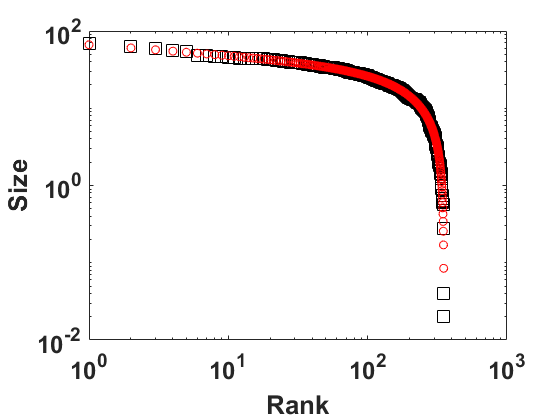}}
	\label{India_margin}
\caption{Plots of actual (Black square) and predicted (Red circle) sizes over the rank 
	for the percentage of vote-margin of Winners in Indian Lok Sabha Elections} 
\end{figure}

\begin{figure}
	\subfloat[Year 2019]{
	\includegraphics[width=0.3\textwidth]{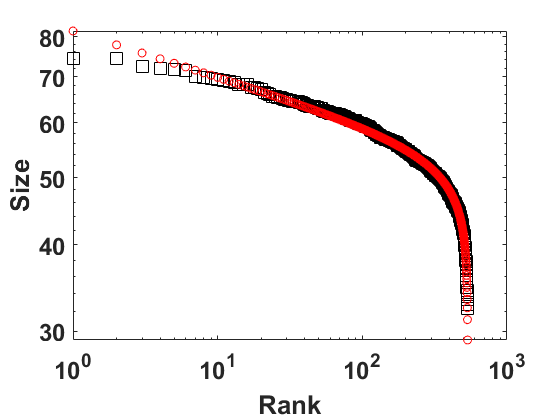}}
	\subfloat[Year 2014]{
	\includegraphics[width=0.3\textwidth]{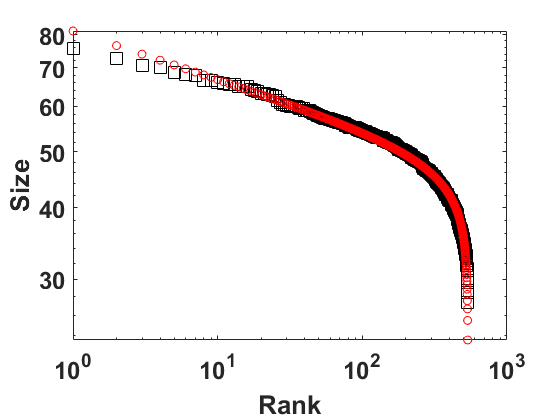}}
	\subfloat[Year 2009]{
	\includegraphics[width=0.3\textwidth]{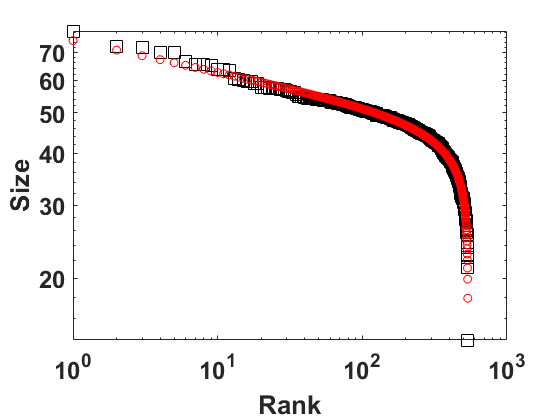}}
\\
	\subfloat[Year 2004]{
	\includegraphics[width=0.3\textwidth]{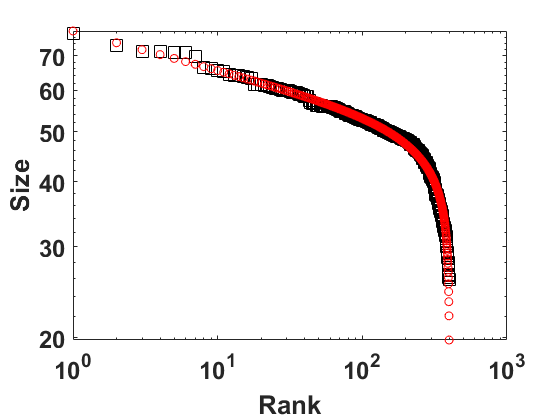}}
	\subfloat[Year 1999]{
	\includegraphics[width=0.3\textwidth]{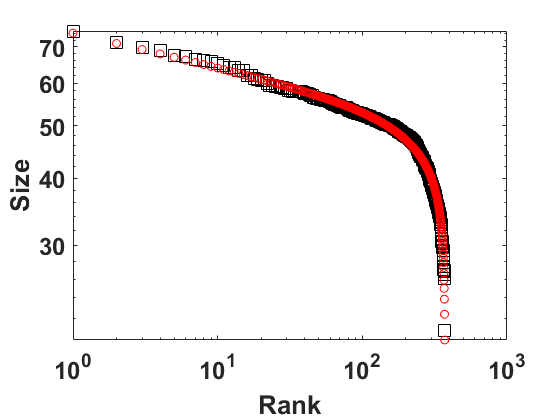}}
	\subfloat[Year 1998]{
	\includegraphics[width=0.3\textwidth]{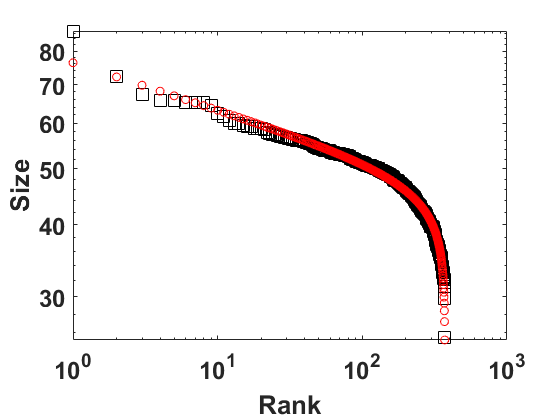}}
\\
	\subfloat[Year 1996]{
	\includegraphics[width=0.3\textwidth]{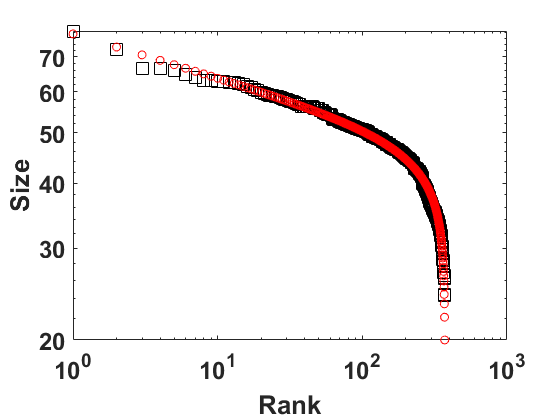}}
	\subfloat[Year 1991]{
	\includegraphics[width=0.3\textwidth]{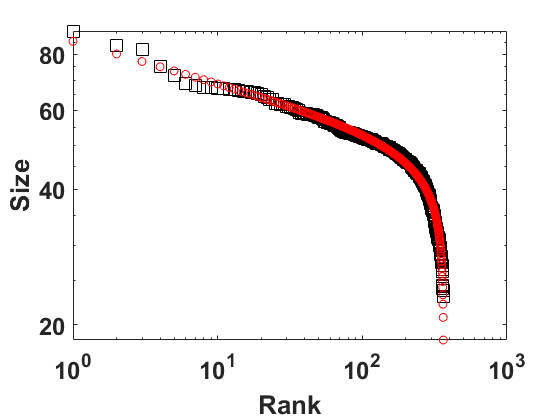}}
	\subfloat[Year 1989]{
	\includegraphics[width=0.3\textwidth]{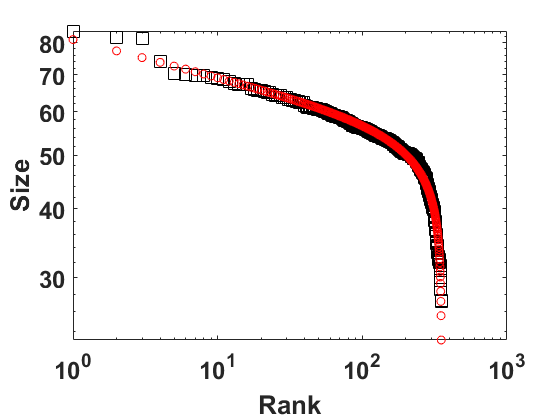}}
\\
	\subfloat[Year 1984]{
	\includegraphics[width=0.3\textwidth]{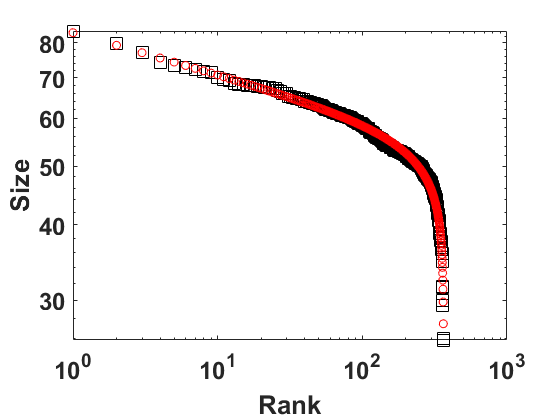}}
	\subfloat[Year 1980]{
	\includegraphics[width=0.3\textwidth]{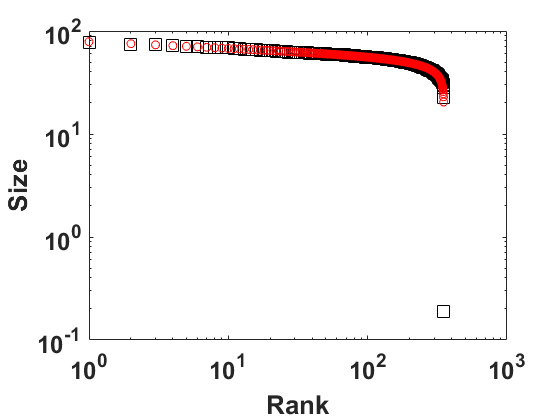}}
	\label{India_votep}
\caption{Plots of actual (Black square) and predicted (Red circle) sizes over the rank 
	for the Vote percentage of Winners in Indian Lok Sabha Elections} 
\end{figure}

\begin{figure}
	\subfloat[Year 2018]{
	\includegraphics[width=0.25\textwidth]{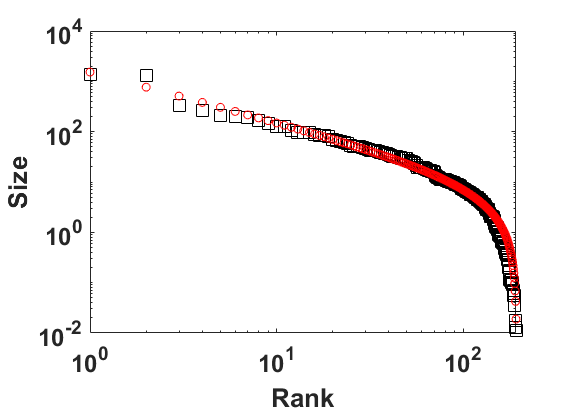}}
	\subfloat[Year 2017]{
	\includegraphics[width=0.25\textwidth]{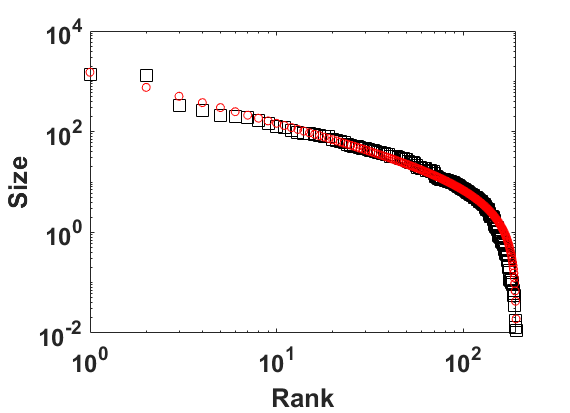}}
	\subfloat[Year 2016]{
	\includegraphics[width=0.25\textwidth]{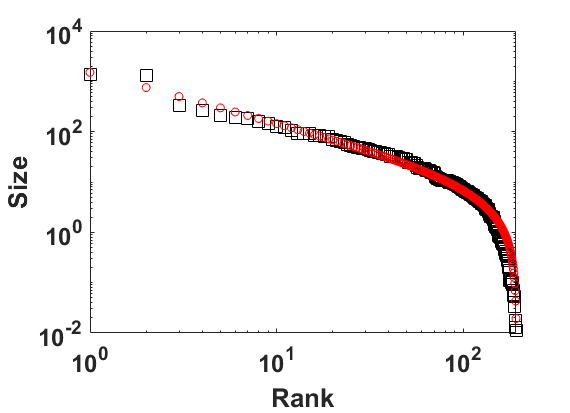}}
	\subfloat[Year 2015]{
	\includegraphics[width=0.25\textwidth]{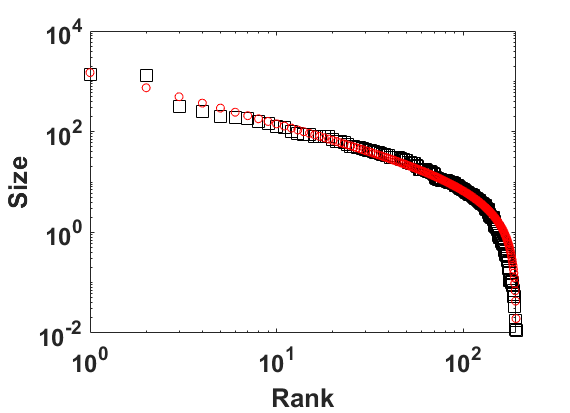}}
\\
	\subfloat[Year 2014]{
	\includegraphics[width=0.25\textwidth]{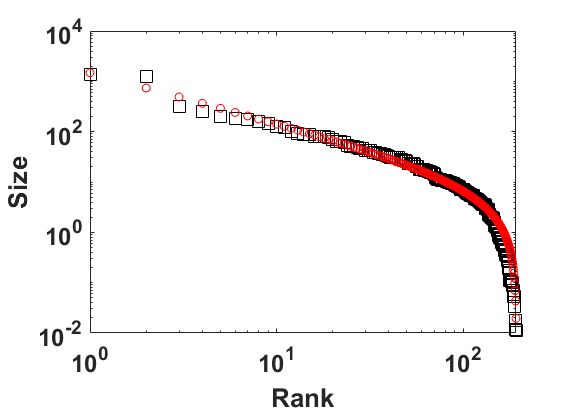}}
	\subfloat[Year 2013]{
	\includegraphics[width=0.25\textwidth]{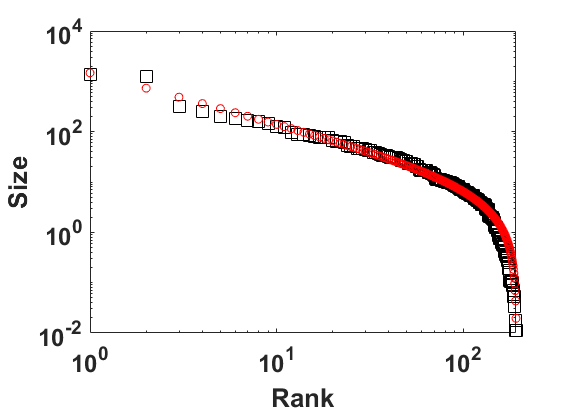}}
	\subfloat[Year 2012]{
	\includegraphics[width=0.25\textwidth]{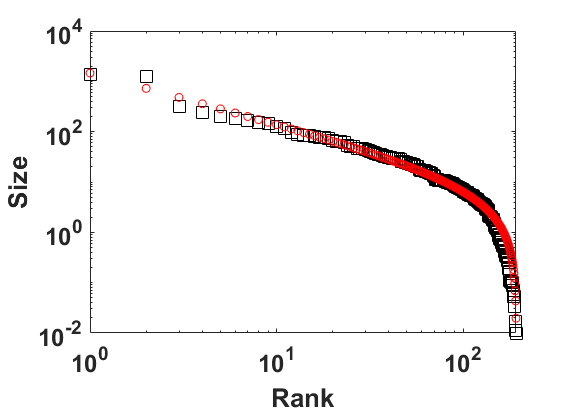}}
	\subfloat[Year 2011]{
	\includegraphics[width=0.25\textwidth]{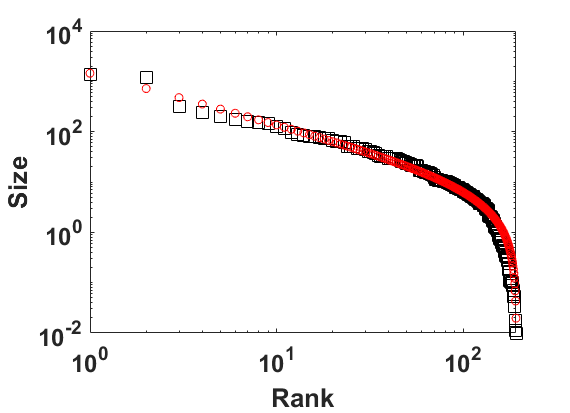}}
\\
	\subfloat[Year 2010]{
	\includegraphics[width=0.25\textwidth]{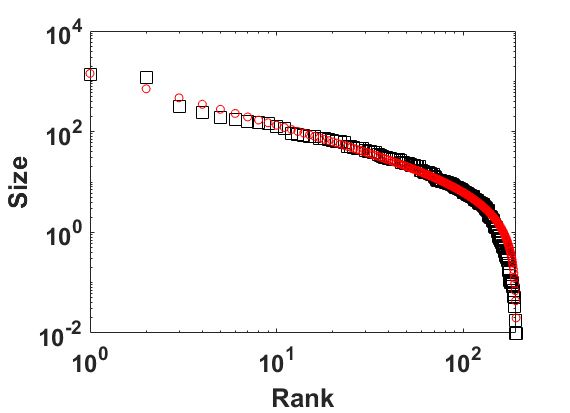}}
	\subfloat[Year 2009]{
	\includegraphics[width=0.25\textwidth]{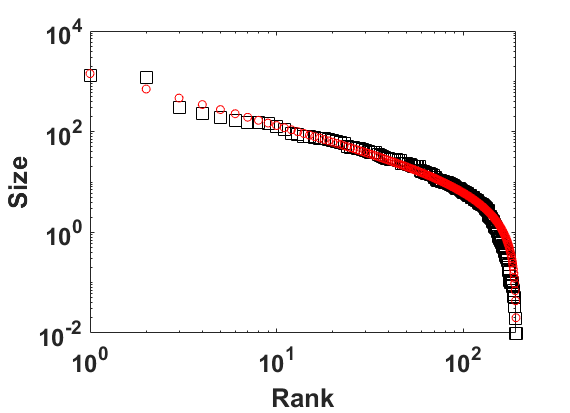}}
	\subfloat[Year 2008]{
	\includegraphics[width=0.25\textwidth]{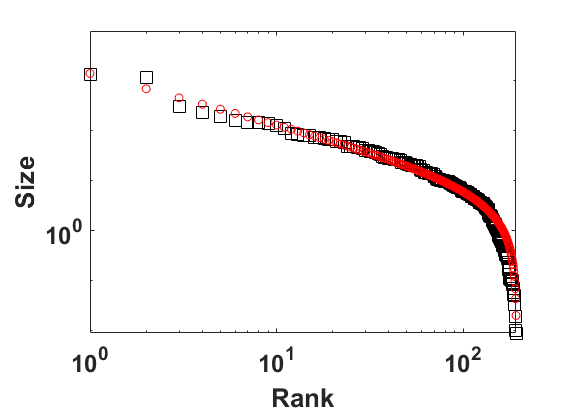}}
	\subfloat[Year 2007]{
	\includegraphics[width=0.25\textwidth]{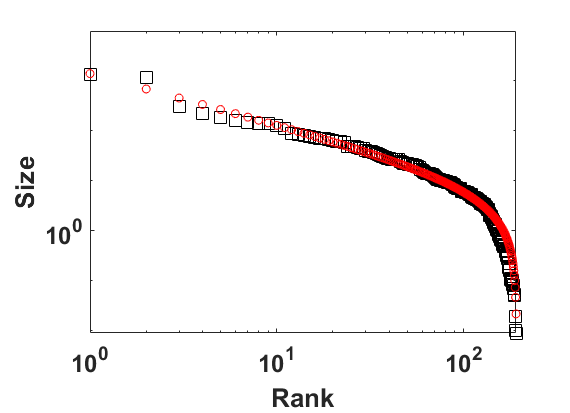}}
\\
	\subfloat[Year 2006]{
	\includegraphics[width=0.25\textwidth]{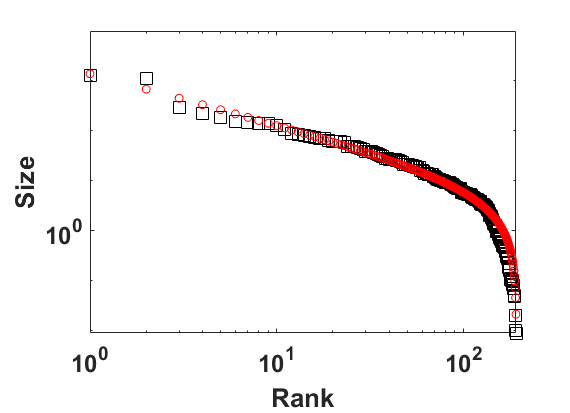}}
	\subfloat[Year 2005]{
	\includegraphics[width=0.25\textwidth]{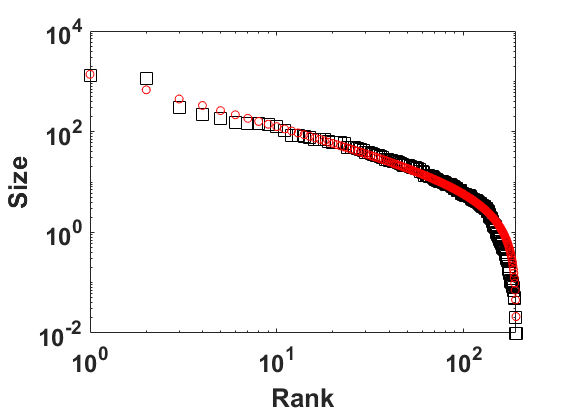}}
	\subfloat[Year 2004]{
	\includegraphics[width=0.25\textwidth]{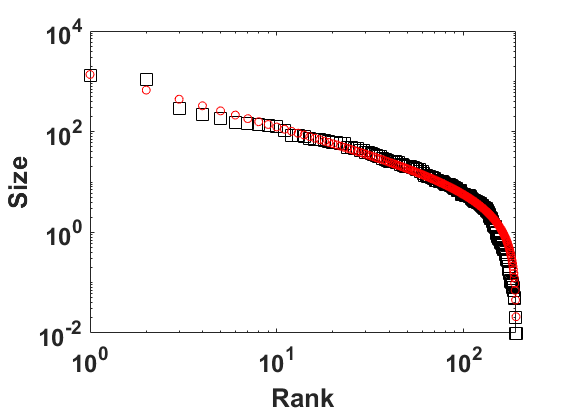}}
	\subfloat[Year 2003]{
	\includegraphics[width=0.25\textwidth]{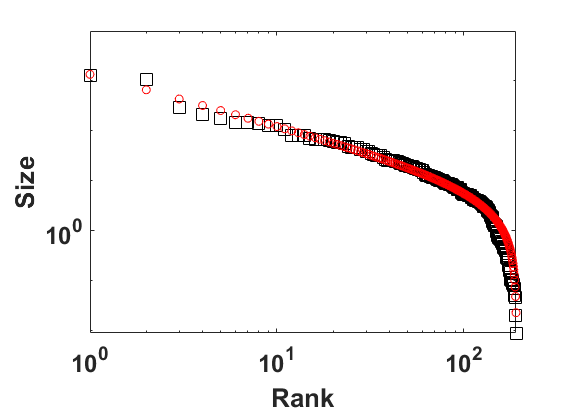}}
\\
	\subfloat[Year 2002]{
	\includegraphics[width=0.25\textwidth]{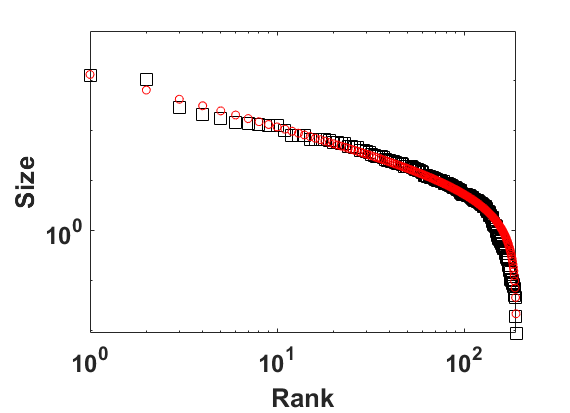}}
	\subfloat[Year 2001]{
	\includegraphics[width=0.25\textwidth]{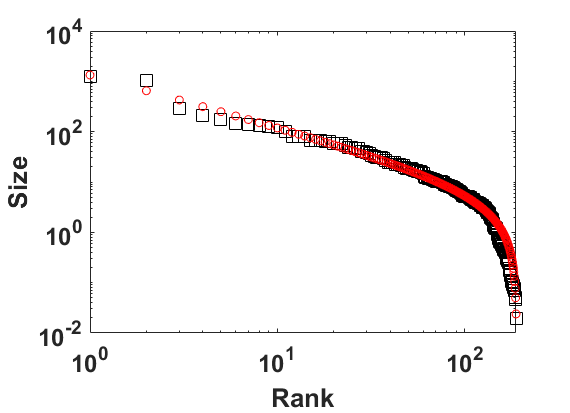}}
	\label{world_pop}
\caption{Plots of actual (Black square) and predicted (Red circle) sizes over the rank 
	for the population of different countries across the world} 
\end{figure}

\begin{figure}
	\subfloat[Year 2016]{
	\includegraphics[width=0.25\textwidth]{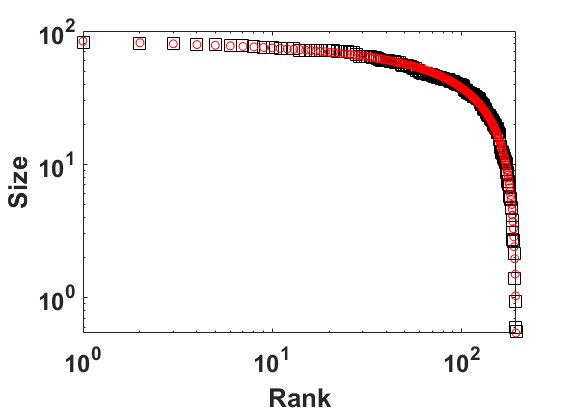}}
	\subfloat[Year 2015]{
	\includegraphics[width=0.25\textwidth]{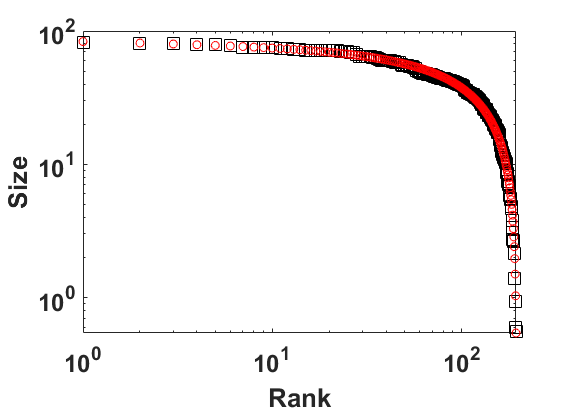}}
	\subfloat[Year 2014]{
	\includegraphics[width=0.25\textwidth]{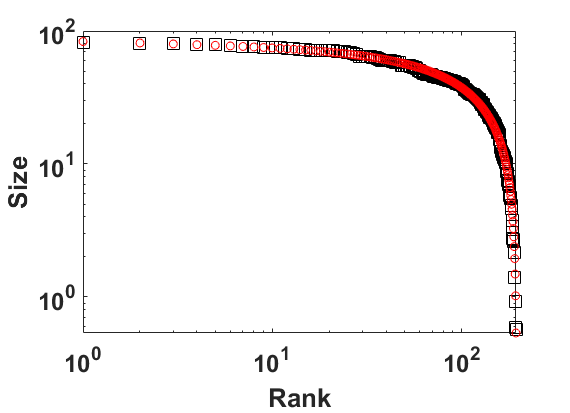}}
	\subfloat[Year 2013]{
	\includegraphics[width=0.25\textwidth]{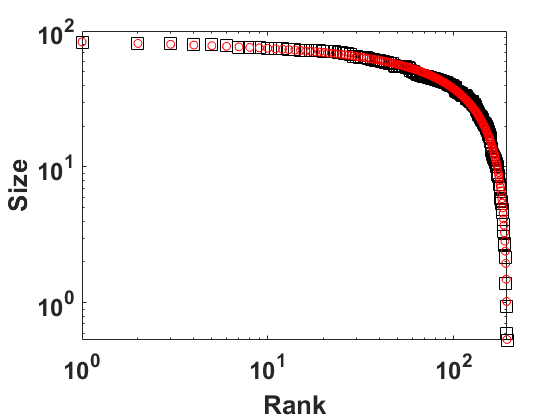}}
\\
	\subfloat[Year 2012]{
	\includegraphics[width=0.25\textwidth]{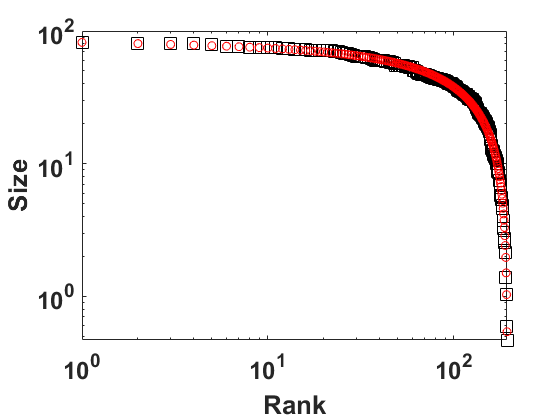}}
	\subfloat[Year 2011]{
	\includegraphics[width=0.25\textwidth]{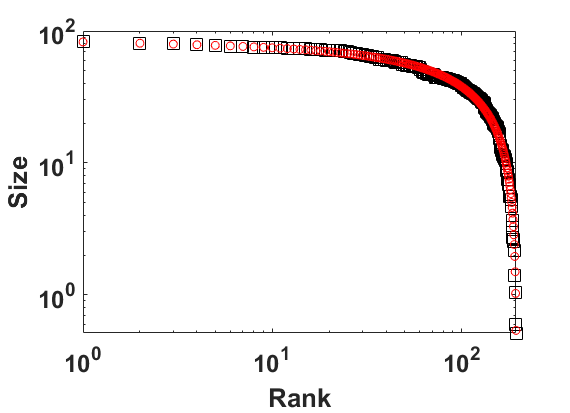}}
	\subfloat[Year 2010]{
	\includegraphics[width=0.25\textwidth]{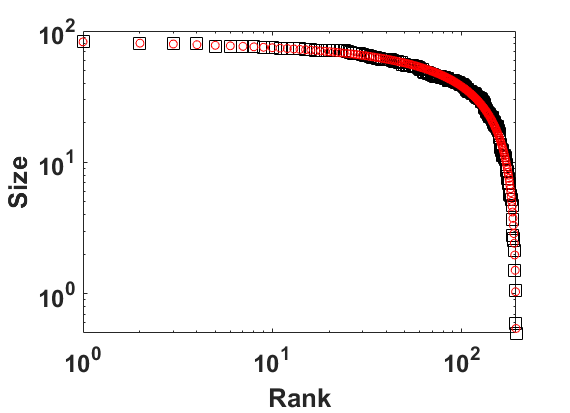}}
	\subfloat[Year 2009]{
	\includegraphics[width=0.25\textwidth]{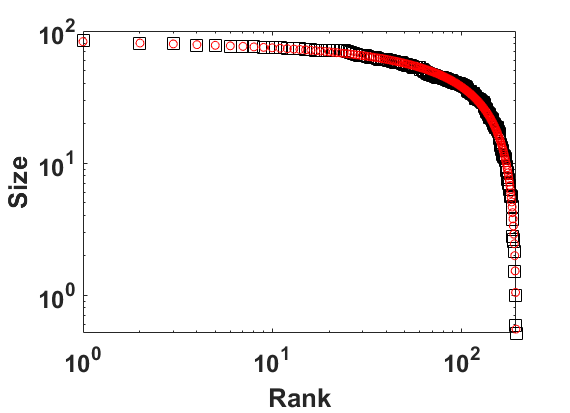}}
\\
	\subfloat[Year 2008]{
	\includegraphics[width=0.25\textwidth]{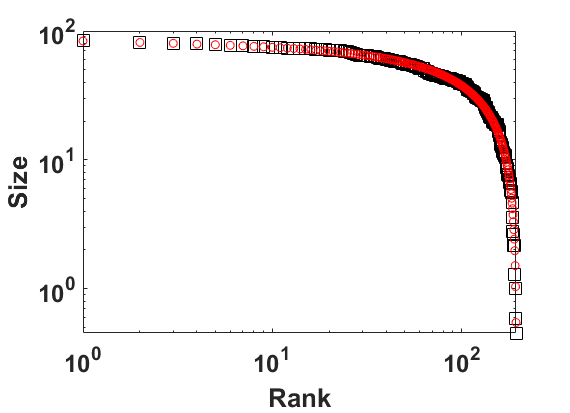}}
	\subfloat[Year 2007]{
	\includegraphics[width=0.25\textwidth]{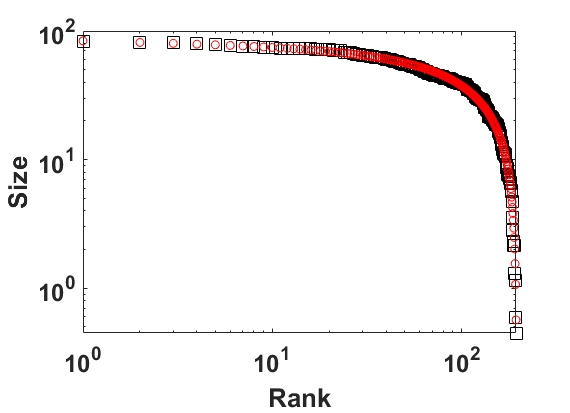}}
	\subfloat[Year 2006]{
	\includegraphics[width=0.25\textwidth]{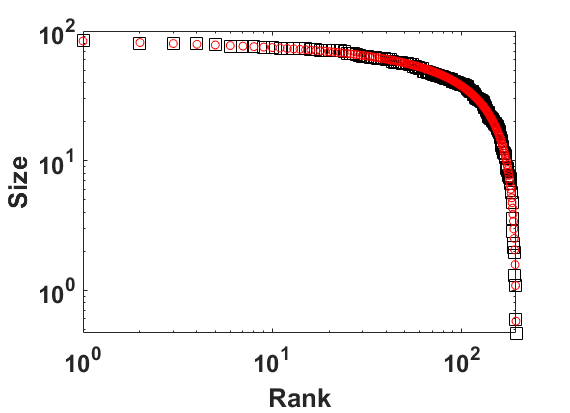}}
	\subfloat[Year 2005]{
	\includegraphics[width=0.25\textwidth]{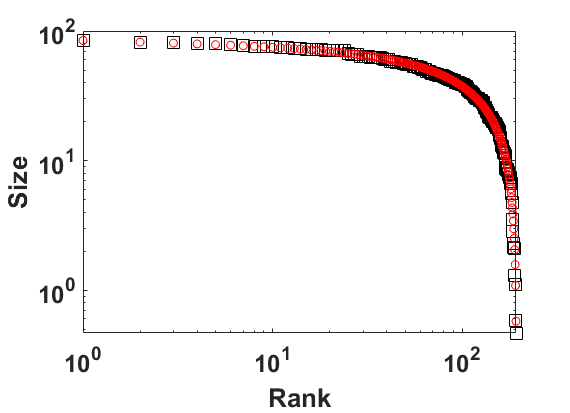}}
\\
	\subfloat[Year 2004]{
	\includegraphics[width=0.25\textwidth]{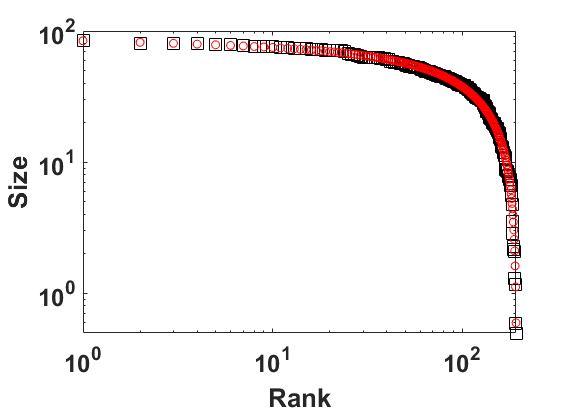}}
	\subfloat[Year 2003]{
	\includegraphics[width=0.25\textwidth]{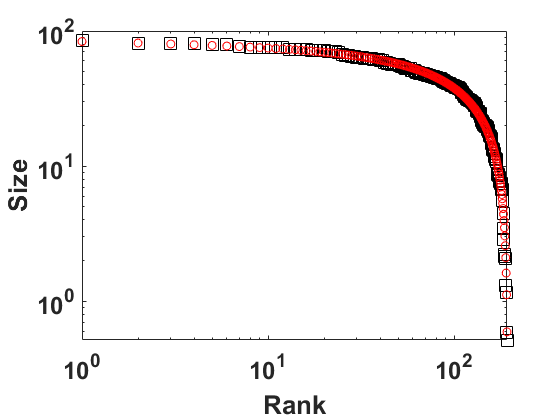}}
	\subfloat[Year 2002]{
	\includegraphics[width=0.25\textwidth]{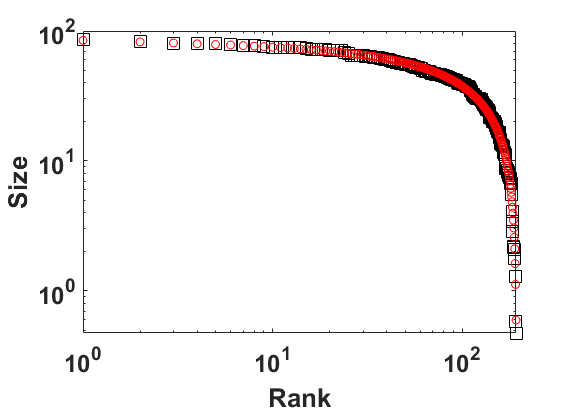}}
	\subfloat[Year 2001]{
	\includegraphics[width=0.25\textwidth]{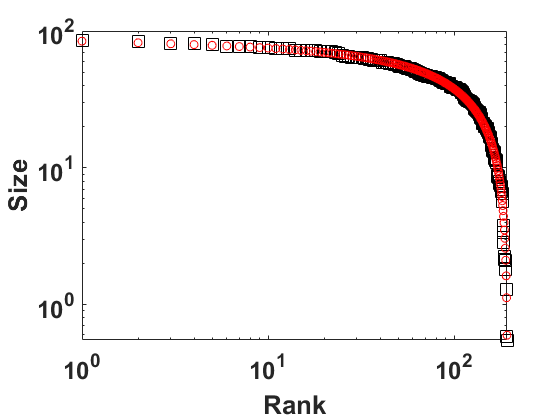}}
	\label{world_ag}
\caption{Plots of actual (Black square) and predicted (Red circle) sizes over the rank 
	for the agricultural land percentage of different countries across the world} 
\end{figure}

\begin{figure}
	\subfloat[Year 2001]{
	\includegraphics[width=0.25\textwidth]{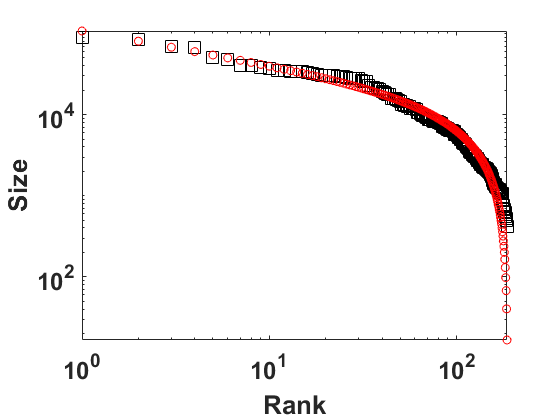}}
	\subfloat[Year 2002]{
	\includegraphics[width=0.25\textwidth]{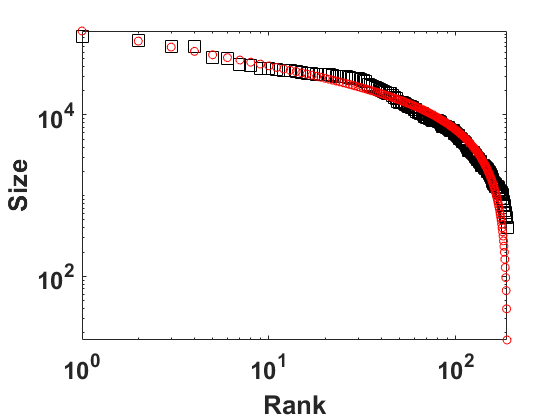}
}
	\subfloat[Year 2003]{
	\includegraphics[width=0.25\textwidth]{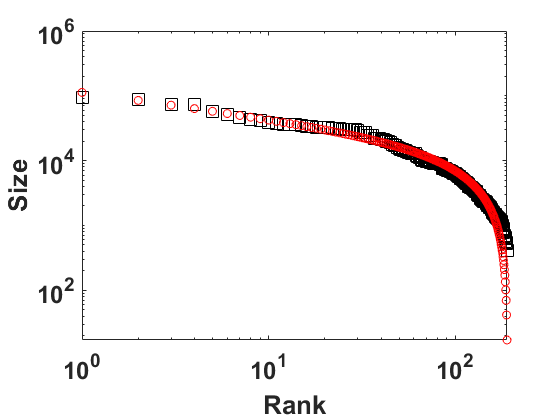}}
	\subfloat[Year 2004]{
	\includegraphics[width=0.25\textwidth]{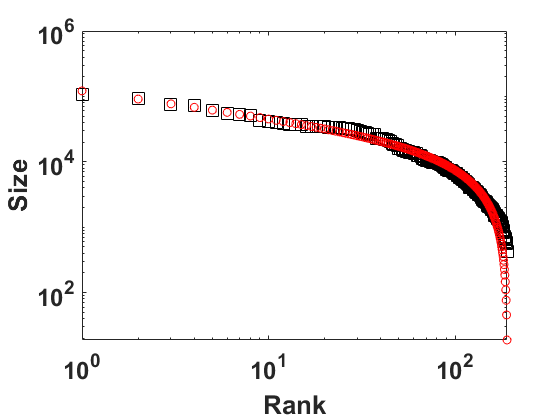}}
\\
	\subfloat[Year 2005]{
	\includegraphics[width=0.25\textwidth]{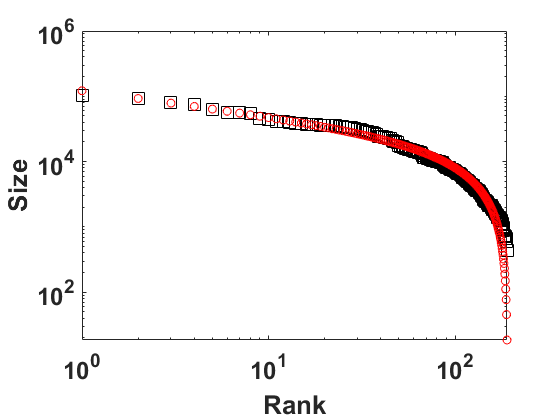}}
	\subfloat[Year 2006]{
	\includegraphics[width=0.25\textwidth]{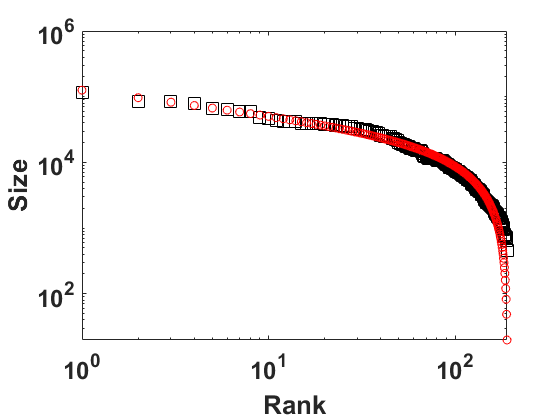}}
	\subfloat[Year 2007]{
	\includegraphics[width=0.25\textwidth]{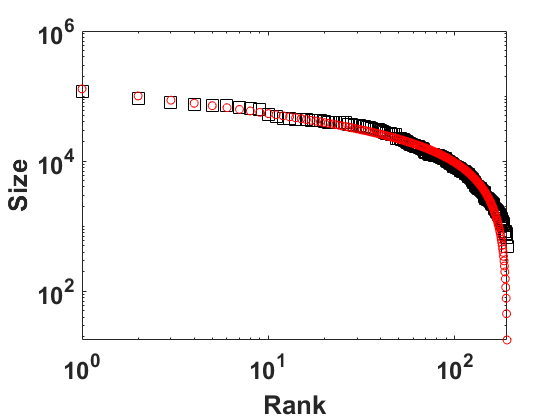}}
	\subfloat[Year 2008]{
	\includegraphics[width=0.25\textwidth]{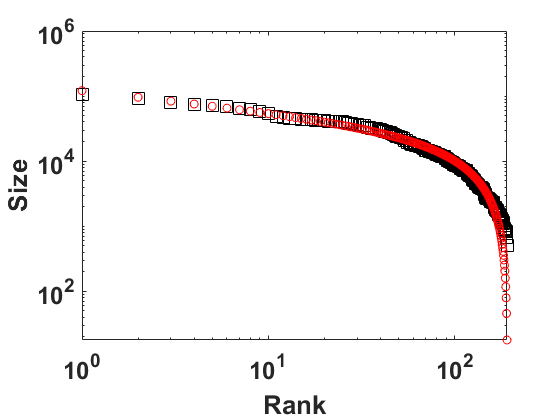}}
\\
	\subfloat[Year 2009]{
	\includegraphics[width=0.25\textwidth]{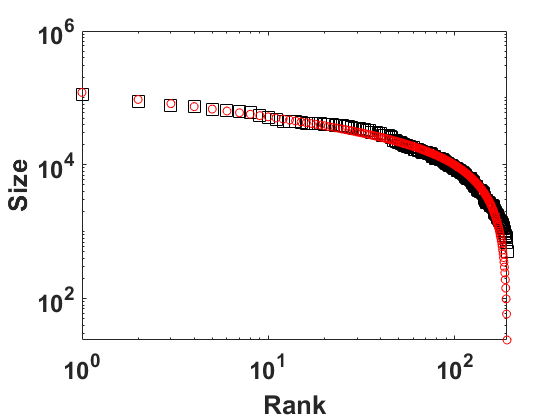}}
	\subfloat[Year 2010]{
	\includegraphics[width=0.25\textwidth]{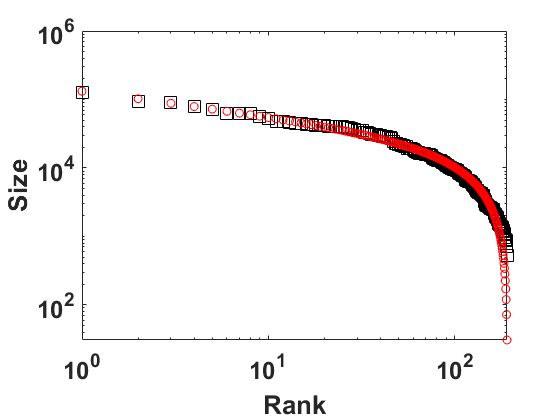}}
	\subfloat[Year 2011]{
	\includegraphics[width=0.25\textwidth]{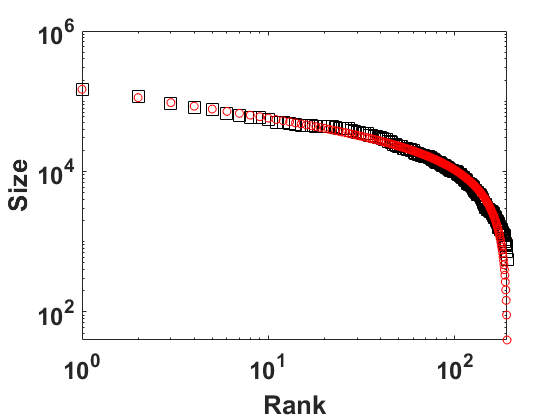}}
	\subfloat[Year 2012]{
	\includegraphics[width=0.25\textwidth]{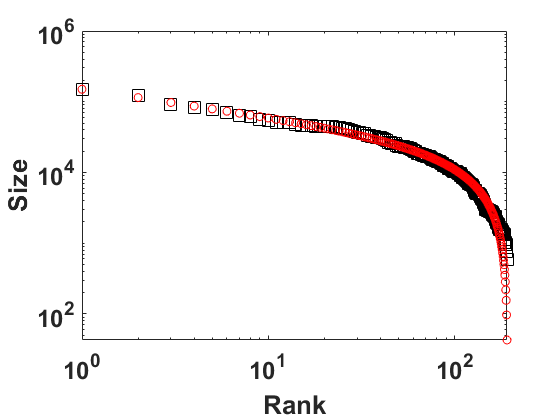}}
\\
	\subfloat[Year 2013]{
	\includegraphics[width=0.25\textwidth]{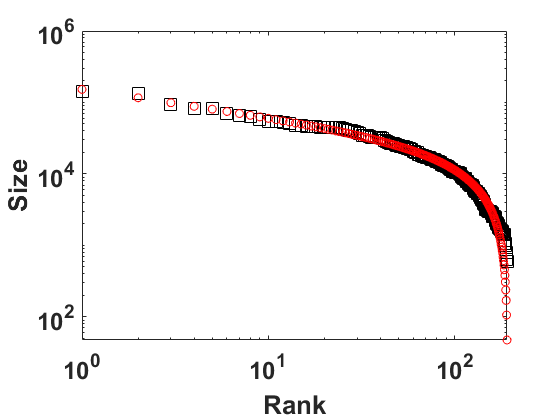}}
	\subfloat[Year 2014]{
	\includegraphics[width=0.25\textwidth]{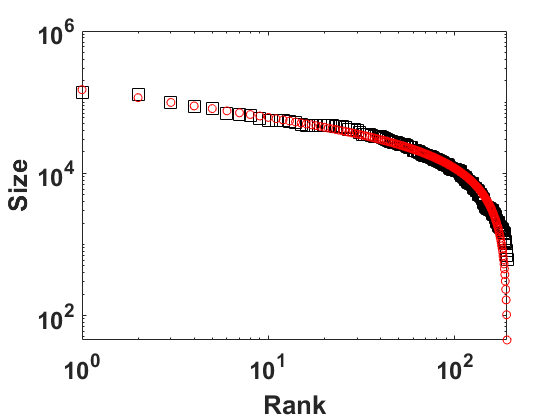}}
	\subfloat[Year 2015]{
	\includegraphics[width=0.25\textwidth]{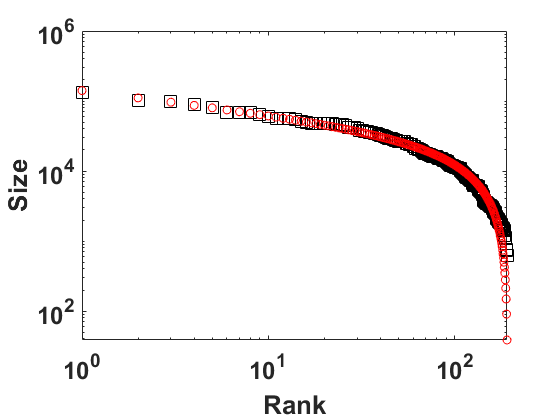}}
	\subfloat[Year 2016]{
	\includegraphics[width=0.25\textwidth]{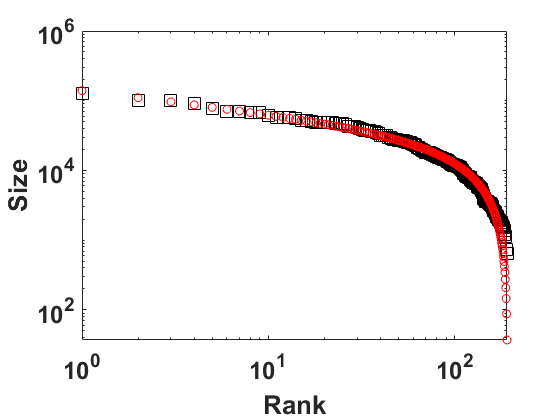}}
\\
	\subfloat[Year 2017]{
	\includegraphics[width=0.25\textwidth]{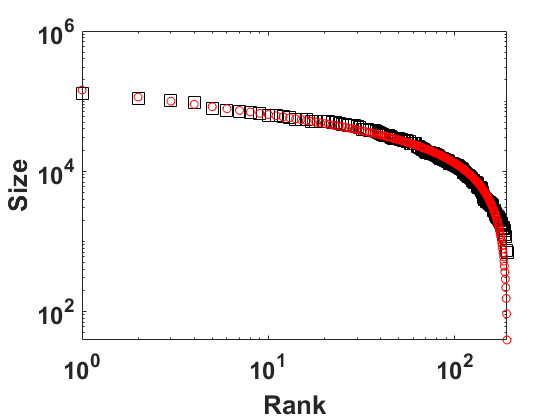}}
	\subfloat[Year 2018]{
	\includegraphics[width=0.25\textwidth]{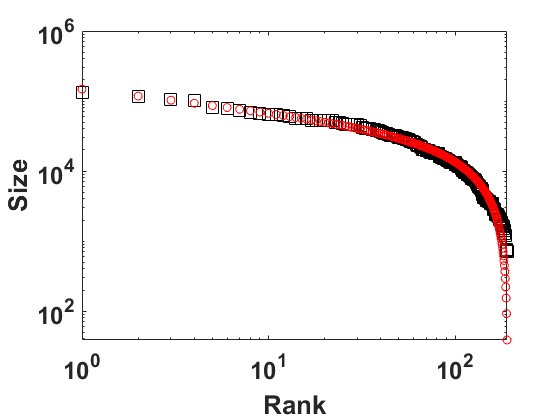}}
	\label{world_gdp}
\caption{Plots of actual (Black square) and predicted (Red circle) sizes over the rank 
	for the per capita GDP of different countries across the world} 
\end{figure}

\begin{figure}
	\subfloat[Year 2016]{
	\includegraphics[width=0.3\textwidth]{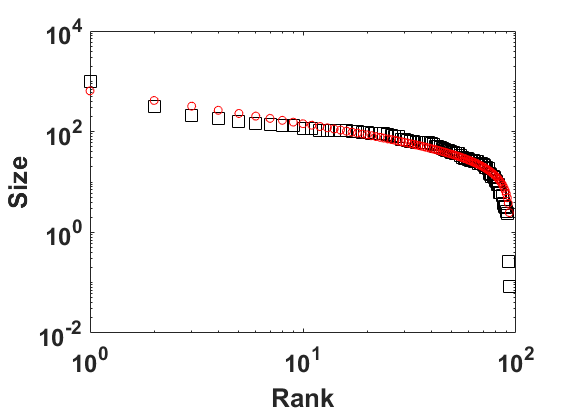}}
	\subfloat[Year 2015]{
	\includegraphics[width=0.3\textwidth]{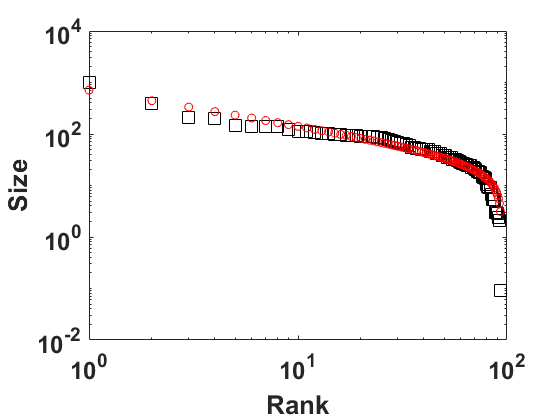}}
	\subfloat[Year 2014]{
	\includegraphics[width=0.3\textwidth]{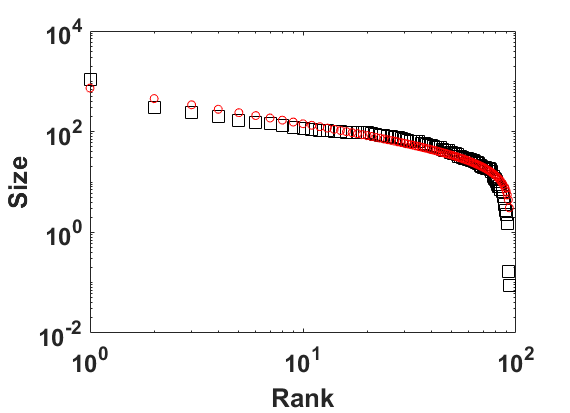}}
\\	\subfloat[Year 2013]{
	\includegraphics[width=0.3\textwidth]{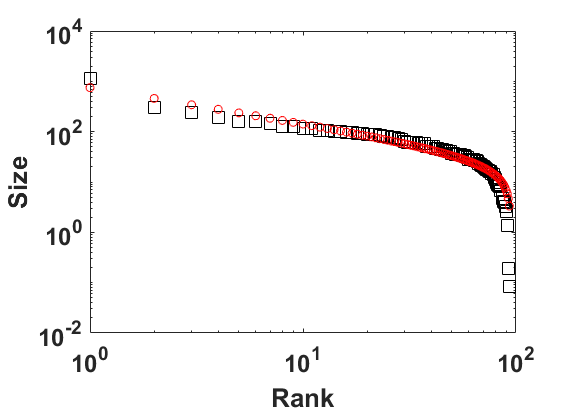}}
	\subfloat[Year 2012]{
	\includegraphics[width=0.3\textwidth]{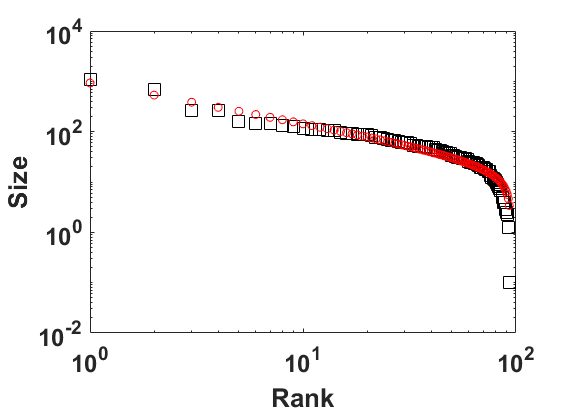}}
	\subfloat[Year 2011]{
	\includegraphics[width=0.3\textwidth]{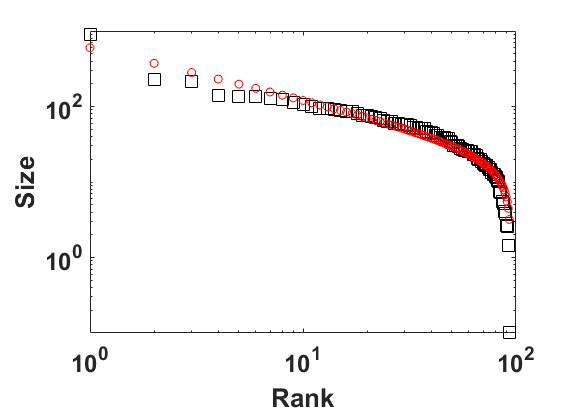}}
\\
	\subfloat[Year 2010]{
	\includegraphics[width=0.3\textwidth]{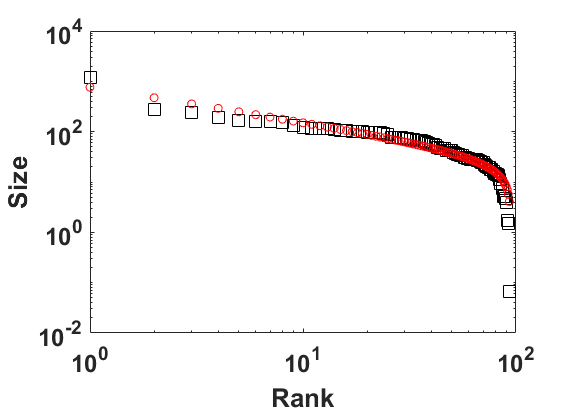}}
	\subfloat[Year 2009]{
	\includegraphics[width=0.3\textwidth]{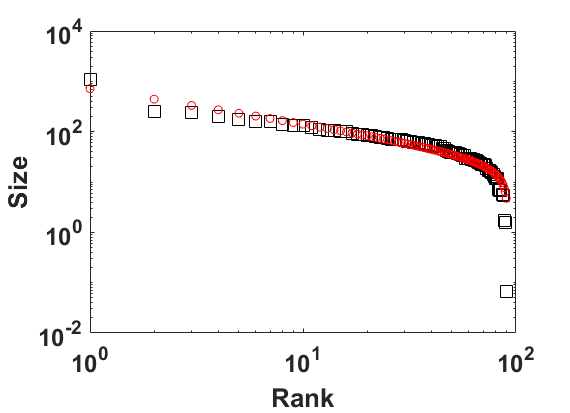}}
	\subfloat[Year 2008]{
	\includegraphics[width=0.3\textwidth]{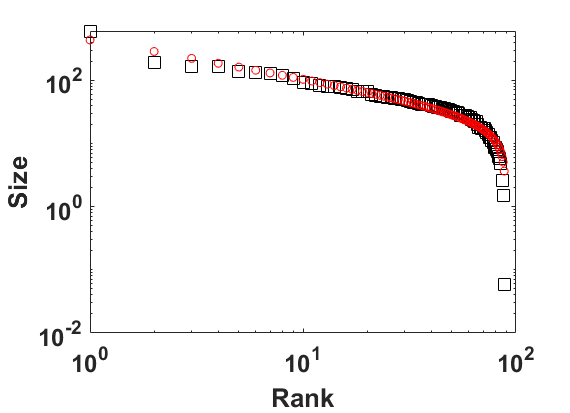}}
\\
	\subfloat[Year 2007]{
	\includegraphics[width=0.3\textwidth]{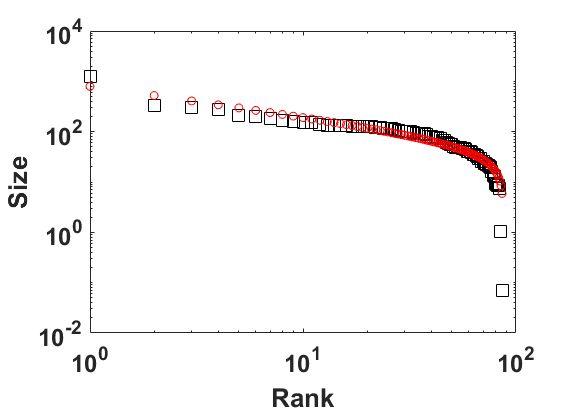}}
	\subfloat[Year 2006]{
	\includegraphics[width=0.30\textwidth]{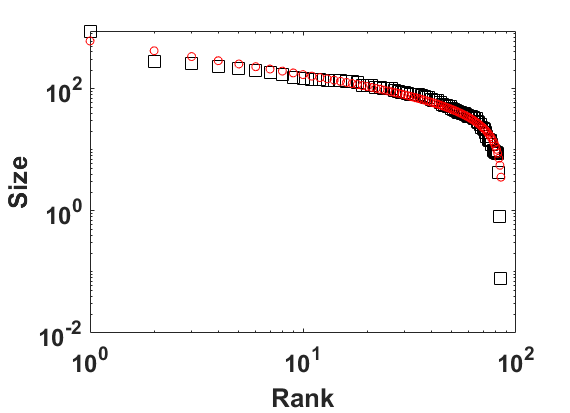}}
	\label{world_BI}
\caption{Plots of actual (Black square) and predicted (Red circle) sizes over the rank 
	for the Buffet indicator of different countries across the world} 
\end{figure}

\end{document}